\documentclass{aa}
\usepackage[varg]{txfonts}
\usepackage{graphicx}
\usepackage{natbib,twoopt}
\usepackage{ulem}
\usepackage{linenoaa}
\linenumbers
\usepackage[breaklinks=true]{hyperref} 
\bibpunct{(}{)}{;}{a}{}{,}             

\makeatletter
  \newcommandtwoopt{\citeads}[3][][]{\href{http://adsabs.harvard.edu/abs/#3}%
    {\def\hyper@linkstart##1##2{}%
     \let\hyper@linkend\@empty\citealp[#1][#2]{#3}}}
  \newcommandtwoopt{\citepads}[3][][]{\href{http://adsabs.harvard.edu/abs/#3}%
    {\def\hyper@linkstart##1##2{}%
     \let\hyper@linkend\@empty\citep[#1][#2]{#3}}}
  \newcommandtwoopt{\citetads}[3][][]{\href{http://adsabs.harvard.edu/abs/#3}%
    {\def\hyper@linkstart##1##2{}%
     \let\hyper@linkend\@empty\citet[#1][#2]{#3}}}
  \newcommandtwoopt{\citeyearads}[3][][]%
    {\href{http://adsabs.harvard.edu/abs/#3}
    {\def\hyper@linkstart##1##2{}%
     \let\hyper@linkend\@empty\citeyear[#1][#2]{#3}}}
\makeatother

\newcommand{\logt}{log$(T$\,[K]$)$}

\begin{document}

\title{Slowly positively drifting bursts generated by large-scale magnetic reconnection}

\author{Alena Zemanov\'a\inst{\ref{inst1}}
\and Marian Karlick\'y\inst{\ref{inst1}}
\and Jaroslav Dud\'{\i}k\inst{\ref{inst1}}
\and Jana Ka\v{s}parov\'a\inst{\ref{inst1}}
\and J\'an Ryb\'ak\inst{\ref{inst2}}}


\institute{Astronomical Institute of the Czech Academy of Sciences, Fri\v{c}ova
298, CZ-25165 Ond\v{r}ejov, Czech Republic \label{inst1} \and Astronomical
Institute, Slovak Academy of Sciences, Tatransk\'{a} Lomnica, 059 60 Vysok\'e Tatry,
Slovakia
\label{inst2}}

\date{Received date /
Accepted date }

\abstract
{The slowly positively drifting bursts (SPDBs) are rarely observed in radio emission of solar flares.} 
{To understand how the SPDBs are generated, we studied the radio observations at 600--5000 MHz together with the imaging observations made in ultraviolet (UV) and extreme ultraviolet (EUV) during the SPDB-rich C8.7 flare of 2014 May 10 (SOL2014-05-10T0702).}
{Because the SPDBs propagate towards locations of higher plasma density, we studied their associations with individual flare kernels, located either within the flare core itself, or distributed at longer distances, but connected to the flaring region by large-scale hot loops. For each kernel we constructed light curves using 1600 \AA~and 304 \AA~observations and compared these light curves with the temporal evolution of radio flux at 1190 MHz, representing all observed groups of SPDBs. We also analysed the UV/EUV observations to understand the evolution of magnetic connectivity during the flare.}
{The flare starts with a growing hot sigmoid observed in 131 \AA. As the sigmoid evolves, it extends to and interacts with a half dome present within the active region. The evolving sigmoid reconnects at the respective hyperbolic flux tube, producing large-scale magnetic connections and an EUV swirl. Three groups of SPDBs are observed during this large-scale magnetic reconnection, along with a group of narrow-band type III bursts. The light curves of a kernel corresponding to the footpoint of spine line analogue show good agreement with the radio flux at 1190 MHz, indicating that the SPDBs are produced by the large-scale magnetic reconnection at the half dome. In addition, one of the kernels appeared in the neighbouring active region and also showed a similar evolution to the radio flux, implying that beams of accelerated particles can synchronize radio and UV/EUV light curves across relatively large distances.} 

\keywords{Sun: flares - Sun: radio radiation - Sun: UV radiation - Sun: X-rays,
gamma rays}

\authorrunning{Zemanov\'a et al.}
\titlerunning{SPDBs generated by Large-Scale Magnetic Reconnection}
\maketitle
\nolinenumbers
\section{Introduction}
Solar flares are explosive phenomena in the solar atmosphere occurring at locations of accumulated magnetic field energy and electric currents. During flares, this energy is rapidly changed into plasma heating, plasma flows, accelerated particles, and emissions in a broad range of electromagnetic waves: from radio; through optical, ultraviolet (UV), extreme ultraviolet (EUV), and X-rays; to gamma rays \citep{Krucker2008,Schrijver2009,Fletcher2011, Aulanier2012_I}.

In the radio range, solar flares are accompanied by many types of radio bursts \citep[see][]{1979itsr.book.....K,1985srph.book.....M,2004psci.book.....A}; 
(for radio bursts in the decimetre range, see catalogues by \citet[][0.3--1.0 GHz]{Guedel1988}, \citet[][1.0--3.0 GHz]{Isliker1994}, or \citet[][0.8--2.0 GHz]{Jiricka2001}). Among them, the most frequent types are type III and IV (continua) bursts. While type III bursts are generated by electron beams via the plasma emission mechanism \citep{Aschwanden2002,Reid2014,Benz2017,ChenB2018}, the continua observed at GHz range are mostly generated by the gyrosynchrotron emission of non-thermal electrons \citep{1979itsr.book.....K,ChenB2020,Kuznetsov2021}.

In the decimetre range, in addition to the type III bursts,  slowly positively drifting bursts (SPDBs) also occur sometimes and have a frequency drift usually less than $\sim$100 MHz s$^{-1}$ \citep{Karlicky2018,Zemanova2020}. In \citet{Karlicky2018} a single SPDB was shown in connection with a falling plasma blob and interpreted by the thermal conduction front. In \citet{Zemanova2020} a group of SPDBs was observed simultaneously with a brightening of huge magnetic rope. It was interpreted by particle beams propagating in the helical magnetic field structure of this rope. However, the origin of SPDBs is still unclear.

Solar flares produce beams of particles, in most cases electron beams. Owing to these beams, flares need not be localized only at the place where they started.  Specifically, particle beams can travel along magnetic field lines from flares to new locations within the flaring active region or even to distant active regions, where they bombard dense atmospheric layers and heat the chromospheric plasma at these locations. Moreover, particle beams can even trigger flaring processes at these locations, if some free energy is present there. In particular, particle beams generate Langmuir waves that, in the current sheet at these locations, increase the resistivity, and thus trigger the reconnection (see \citealt{Karlicky1989}). 

There is some correlation between  decimetre radio emission and hard X-rays (HRX). For example, \citet{Aschwanden1985} compared the occurrence of decimetre radio bursts (300--1000 MHz) with simultaneous emission in HXR and found an association of 51\%. He also found that 68\% of decimetre radio events usually started at the rise time of HXR, so they considered them to be a phenomenon connected to the impulsive phase of flares.

Solar radio spectra at decimetre wavelengths provide us with information about the processes occurring during solar flares and eruptions; however, they do not show the  positions of the radio sources. Their positions are likely intimately connected to the magnetic field configuration of flaring active regions and their inter-connections; therefore, localizing them may help to understand how they are generated. Spectral-imaging instruments providing observations at decimetre wavelengths are still rare \citep{Tan2023}. To date, the best imaging observation of active and flaring regions in the solar corona is due to the magnetic field being visualized by hot plasma in EUV and X-rays. 

The present paper combines temporal changes in the connectivity of magnetic loops as seen in UV/EUV images, observations of radio bursts in decimetre frequency range, and the light curves (LCs)
made from selected flare kernels to find any association of radio bursts with processes forming distinct flare kernels. Using the models of three-dimensional (3D) magnetic reconnection \citep{Aulanier2006,Pontin2013,Wyper2016a,Wyper2016b}, we disentangle flare processes in the flare active region. In connection with this flare, we recognized a flaring activity even in a neighbouring active region. Then, based on these results, we interpret SPDBs as those generated by particle beams propagating along magnetic field lines that are formed during the 3D magnetic reconnection at the hyperbolic flux tube (HFT). Finally, we discuss their low positive frequency drift and propose its explanation.

\section{Data}
\label{sec_data} 

For this paper we studied radio spectra obtained by the radio spectrographs RT4 and RT5 at Ond\v{r}ejov Observatory of Astronomical Institute of the Czech Academy of Sciences. The spectrograph RT4 works in the frequency range 2000--5000 MHz with the frequency resolution 11.7 MHz and time resolution 10 ms \citep{Jiricka1993,Jiricka2008}. The RT5 works in the 800--2000 MHz range with the frequency resolution 4.7 MHz and time resolution of 10 ms \citep{Jiricka1993,Jiricka2008}. To extend the frequency range down to 600 MHz, we also used one spectrum from the BLEN7M Callisto \citep{Benz2005b}.

The study of radio spectra is supplemented with imaging data obtained by the Atmospheric Imaging Assembly (AIA) \citep{Lemen2012} on board the Solar Dynamic Observatory (SDO) \citep{Pesnell2012}. The AIA provides multiple simultaneous images of the solar disc with 12 s cadence and spatial resolution of 1.5\arcsec. It consists of four identical telescopes working together to provide images of the Sun in seven narrow-band extreme ultraviolet (EUV) filters and two ultraviolet (UV) filters. The EUV images were deconvolved with a point spread function using {\it aia\_deconvolve\_richardsonlucy.pro} and then processed via {\it aia\_prep.pro}, using the standard SolarSoft tree.\footnote[1]{\url{https://www.lmsal.com/solarsoft/ssw\_install.html}} In this study we used images in the  131 \AA, 171 \AA, 193 \AA, 211 \AA, and 304 \AA~EUV filters and in the  1600 \AA~UV filter. In the  SDO/AIA 131 \AA~filter hot emission from \ion{Fe}{XXI} at \logt\,=\,7.05 dominates during the flare, otherwise the filter contains emission from the quiet corona of \ion{Fe}{VIII} at \logt = 5.6; the 171 \AA~filter contains mainly emission produced by \ion{Fe}{IX} at \logt = 5.85, but in a flare the continuum contributes significantly. In 193 \AA~the main contribution comes from \ion{Fe}{XII} at \logt = 6.2, but during the flare \ion{Fe}{XXIV} emission at \logt = 7.25 it can become important. The 211 \AA~passband contains the main contribution from \ion{Fe}{XIV} at \logt = 6.3, but during a flare the continuum enhances, and emission from \ion{Fe}{XVII} (\logt = 6.6) and \ion{Ca}{XVII} (\logt = 6.7) may also contribute. Finally, the 304 \AA~filter covers emission of \ion{He}{II} at \logt = 4.7, while in flares the  hot emission from \ion{Ca}{XVIII} at \logt = 6.85 can contribute (for details see \cite{ODwyer2010}. For the analysis of LCs from selected flare kernels, we used the  SDO/AIA 1600 \AA~UV and 304 \AA~EUV filters. To visualize the coronal loop connection between the flaring active region (AR 12 056) and its neighbourhood, we used SDO/AIA images in 131 \AA~and 193 \AA~filters and applied to them a method of multi-scale Gaussian normalization introduced by \citet{Morgan2014}.

To show the spatial distribution of the magnetic field in the flaring active region and its neighbourhood, we used the longitudinal magnetograms obtained from the Helioseismic and Magnetic Imager (HMI) \citep{Scherrer2012} on  board the  SDO. They were processed by the standard SolarSoft routine {\it aia\_prep.pro} which re-scales them to the AIA pixel size.
As supplementary imaging data, we also used H$\alpha$ solar synoptic images which were provided by the Kanzelhohe Observatory for Solar and Environmental Research (KSO), University of Graz, Austria.

The time evolution of soft X-ray (SXR) flux during the flare was obtained by X-ray Sensors (XRS) on board the Geostationary Operational Environmental Satellite 15 (GOES-15).\footnote[2]{\url{https://www.ncei.noaa.gov/data/goes-space-environment-monitor/access/science/xrs/GOES\_1-15\_XRS\_Science-Quality\_Data\_Readme.pdf}} The XRS provides solar X-ray irradiances for two wavelength bands: 0.5--4.0 \AA~and 1--8 \AA. Here we used the operational data provided by the  Space Weather Prediction Center (SWPC).\footnote[3]{\url{https://ngdc.noaa.gov/stp/satellite/goes/doc/GOES\_XRS\_readme.pdf}}

To complement the observed radio spectra and flare evolution in EUV we used Hard X-ray (HXR) data obtained by the Reuven Ramaty High-Energy Solar Spectroscopic imager (RHESSI) \citep{Lin2002}. The RHESSI images used in this study were taken from the  Flare Image Archive, which is a part of RHESSI Misson Archive.\footnote[4]{\url{https://hesperia.gsfc.nasa.gov/rhessi/mission-archive}} We took images reconstructed using the  CLEAN algorithm and detectors 5 to 9.

%
\begin{figure}
\resizebox{\hsize}{!}{\includegraphics{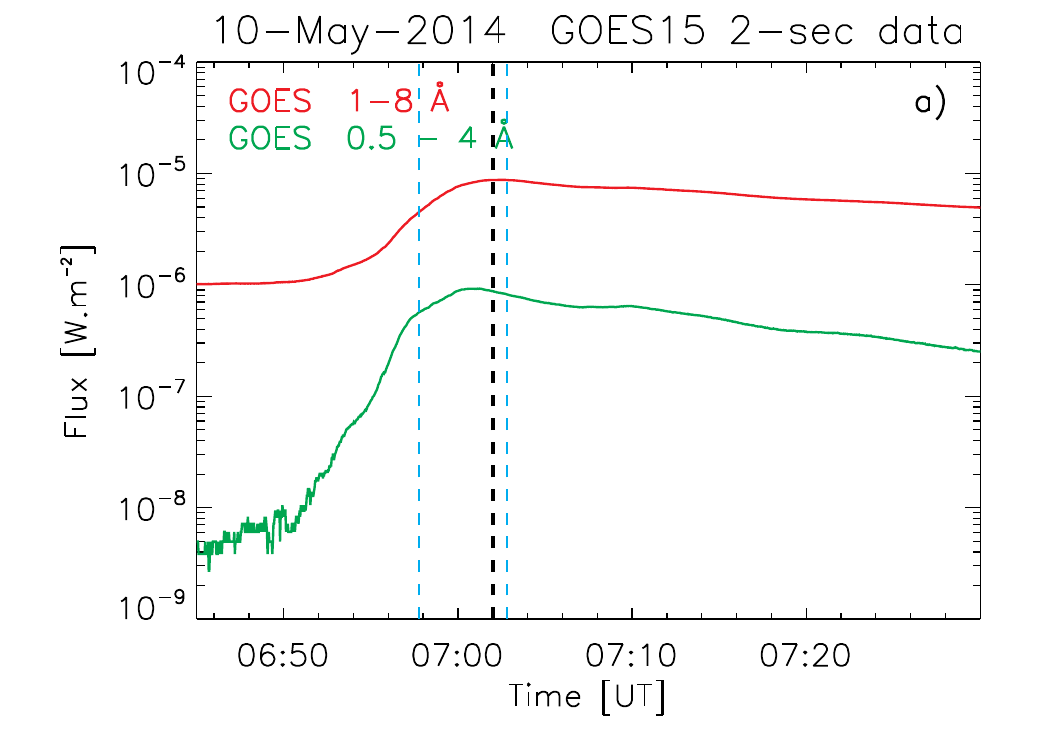}}
\resizebox{\hsize}{!}{\includegraphics{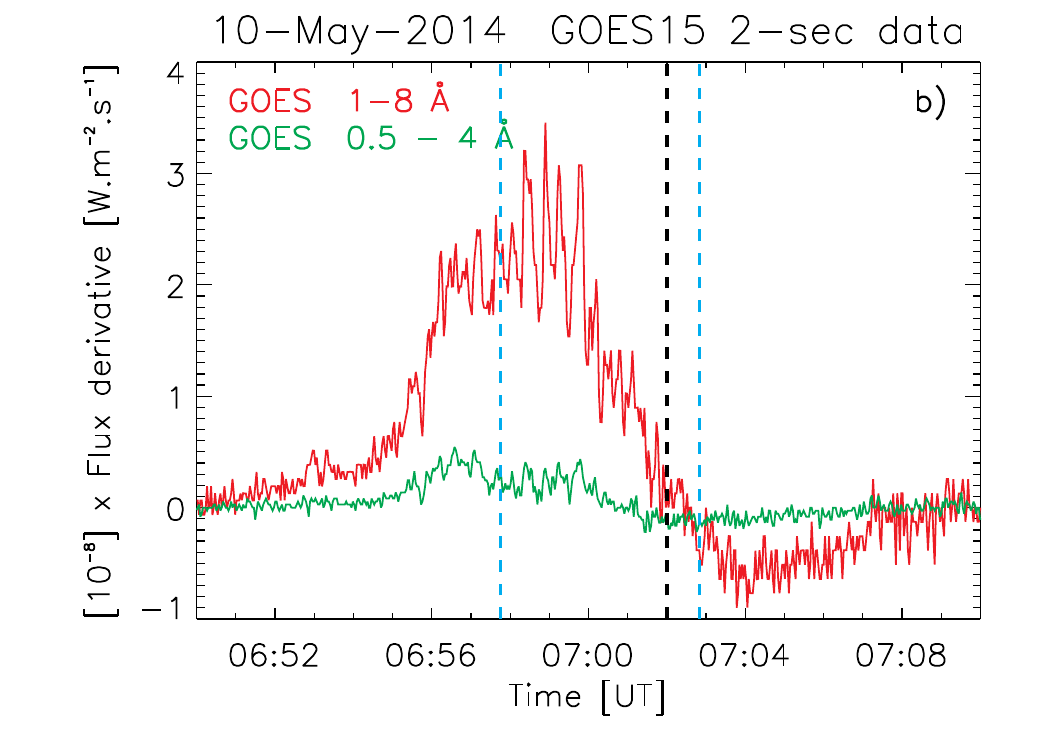}}
\caption{GOES\,15 light curves. (a) Flux in 1-8 and 0.5-4 \AA~during the time interval
06:45-07:30 UT and (b) flux derivative during the time interval 06:45 - 07:15 UT.
Blue vertical dashed lines show the time interval when radio bursts were observed and the black
line marks the flare maximum in SXR.}
    \label{fig_X-ray}
\end{figure}

\section{Observations}
\subsection{The 2014 May 10 flare}
\label{sec_flare}

On 2014 May 10, the Ond\v{r}ejov radiospectrographs observed three groups of SPDBs (Sect.~\ref{sec_radio}) during 06:57--07:03 UT. A summary of the most important radio bursts is given in Table \ref{table:1}. This radio activity occurred during a confined flare of GOES soft X-ray class C8.7 (Fig.~\ref{fig_X-ray}). We note that the occurrence of three distinct groups of SPDBs is unusual, given that these bursts are rare, and we regard this flare as SPDB-rich. The flare itself lasted during 06:51--07:30 UT, with a peak at 07:02 UT (Fig.~\ref{fig_X-ray}a). The derivative of the GOES X-ray flux is shown in panel (b) of Fig.~\ref{fig_X-ray}, and the blue dashed lines mark the time interval 06:57--07:03 UT when groups of SPDBs were observed. Almost all detected radio bursts appeared during its impulsive phase, except  group SPDBs-3, which started a few seconds after the maximum of SXR flux. This makes the flare an excellent candidate for association with observed radio burst activity. Simultaneously with the  SXR C8.7 flare, the  H$\alpha$ flare was also detected during 06:50--08:09 UT in active region NOAA 12\,056. The evolution of the flare is studied in Sect. \ref{sec_euv}.

We note that the GOES satellite sees SXR flux from the whole solar disc, and in principle, any other brightening could occur at a similar time as the C8.7 flare. Therefore, we checked for other solar activity within the solar  disc. In the SDO/AIA 131~\AA~filter we found an enhancement in hot emission within the neighbouring active region NOAA 12\,055. An SXR flare in active region NOAA 12\,055 identified by GOES was observed more than one hour later. The enhancement of hot emission at about 06:57 UT is  studied further in Sect. \ref{sec_uv_lc}.  
%
\begin{figure}[ht!]
\resizebox{\hsize}{!}{\includegraphics{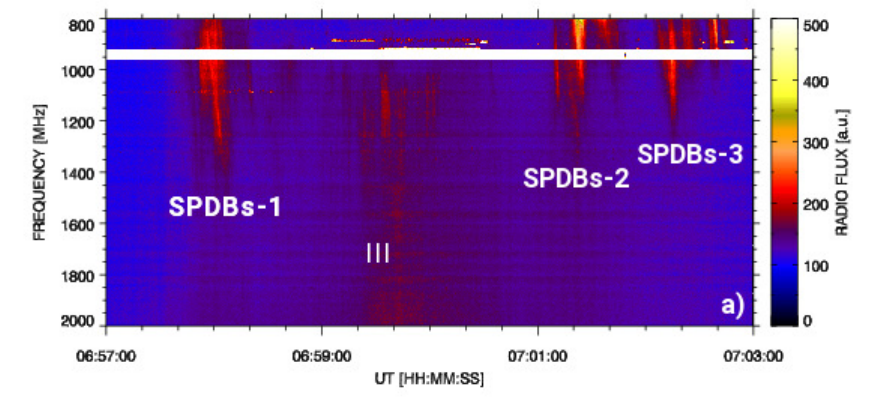}}

\resizebox{\hsize}{!}{\includegraphics{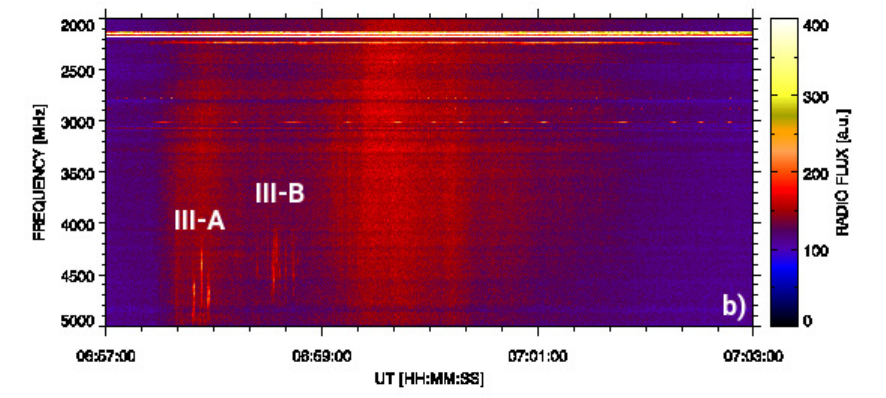}}

\includegraphics[width=8.75cm,clip, viewport=20 25 525 325]{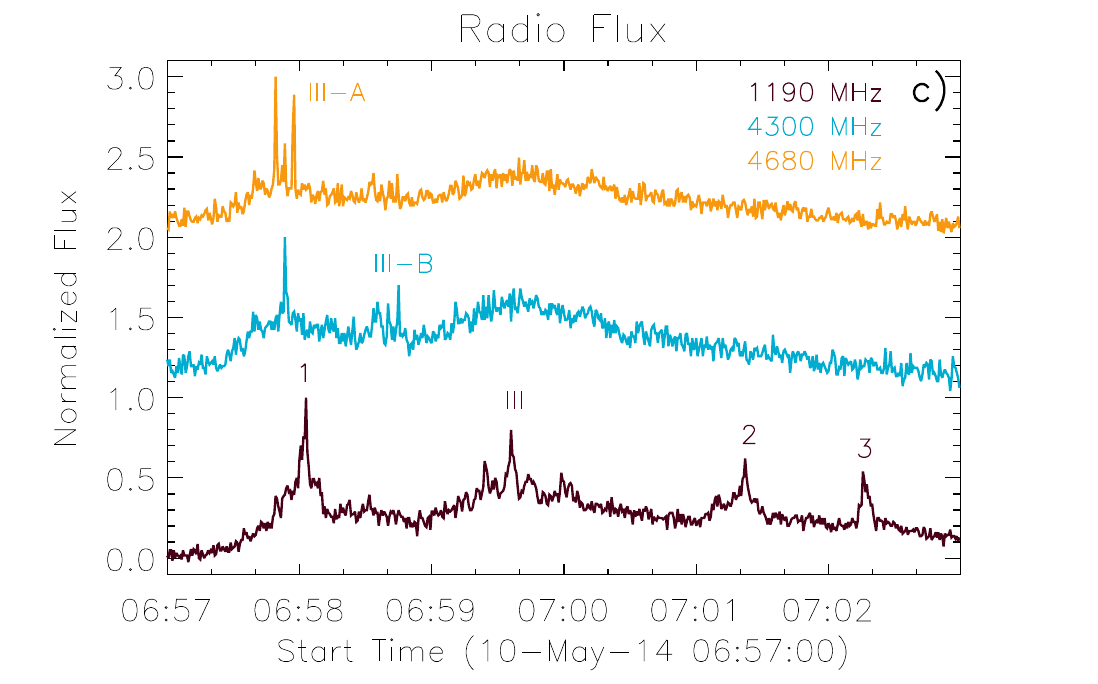}

\includegraphics[width=8.25cm,clip, viewport=8 0 566 450]{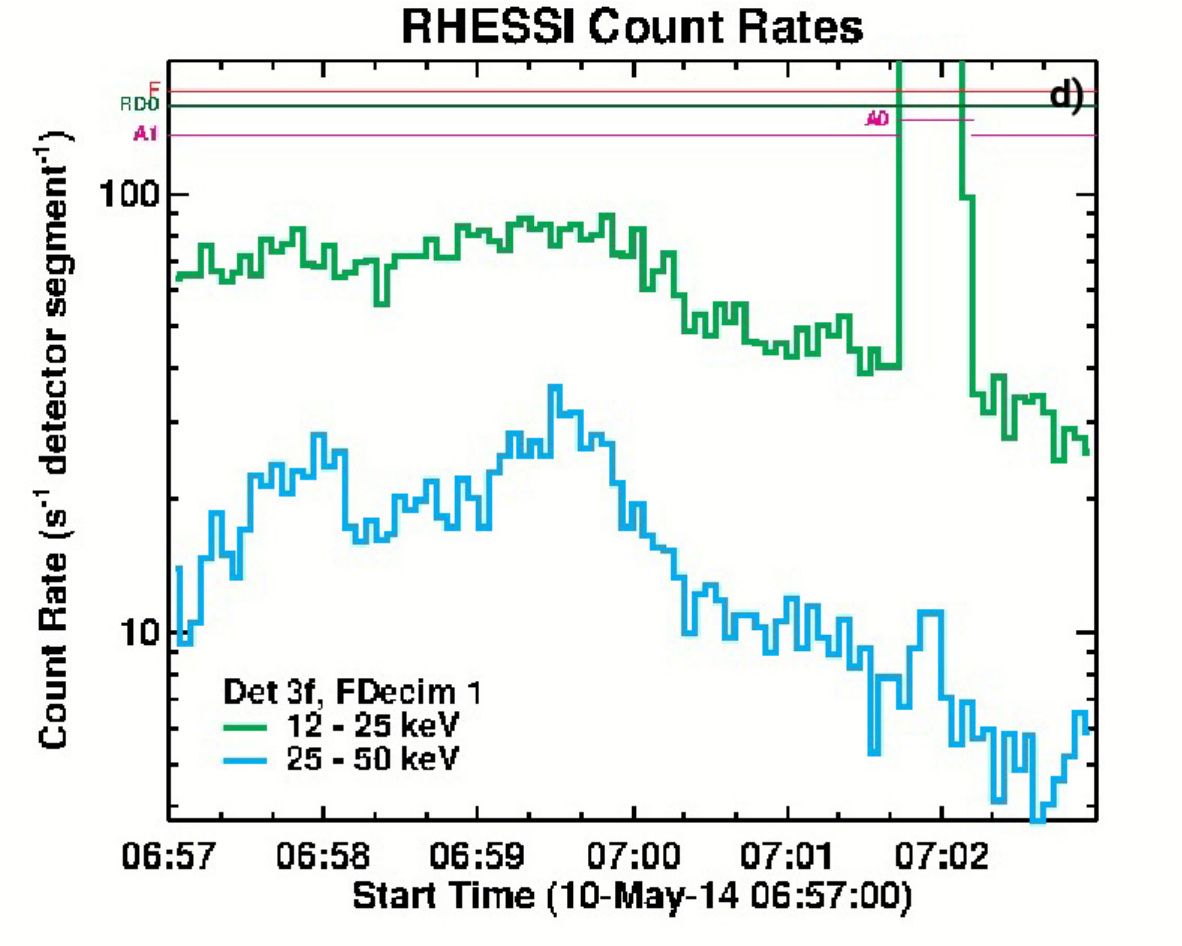}

\caption{Spectra from the Ond\v{r}ejov radiospectrographs and RHESSI HXR light curves.
The radio spectrum in 800--2000 MHz (RT5) (a), in 2000--5000 MHz (RT4) (b), cuts through spectra at 1190, 4300, 4680 MHz (c).
Each cut was normalized to its maximum flux and, for clarity, plotted by adding a constant (0, 1.0, 2.0, respectively).
Individual groups of SPDBs are labelled  1, 2, and  3 and type III bursts   are labelled   III, III-A, III-B;
RHESSI HXR light curves for two energy channels and detector 3: 12-25 keV in green and
25-50 keV in blue (d). The sudden jump in HXR rates at about 07:02 UT is due to attenuator change.}
    \label{fig_rspec}
\end{figure}

\subsection{Radio}
\label{sec_radio}

The radio spectrum within the time interval 06:57--07:03 UT observed during the 2014 May 10 flare (SOL2014-05-10T0702 C8.7) by the Ond\v{r}ejov radiospectrographs RT5 and RT4 is depicted in Fig.~\ref{fig_rspec}a and b, respectively.
Figure~\ref{fig_rspec}a shows the spectrum in the frequency range of 800--2000 MHz. During the time interval 06:57:48--06:58:08 UT, we observed the first group of slowly positively drifting bursts (SPDBs-1). According to the BLEN7M Callisto spectrum, some of the bursts started at 700 MHz. These bursts resembled reverse type III bursts, but their frequency drift, 54--90 MHz s$^{-1}$, was much smaller
than that typical of type III bursts, $\sim$1 GHz s$^{-1}$, in this frequency range \citep{Aschwanden2002}. Considering the plasma emission mechanism for these bursts, an agent driving them was supposed to move through a plasma with a density of about 6--20 $\times$10$^9$~cm$^{-3}$ (fundamental) or 1.5--5 $\times$ 10$^9$~cm$^{-3}$ (harmonic). Further, within 06:59:00--07:00:20 UT, narrow-band type III bursts were observed over frequencies 900--1400 MHz. They were of lower intensity than SDPBs-1 and were accompanied by an enhanced continuum (Fig.~\ref{fig_rspec}a--c). Then, during 07:01:00--07:03:00 UT we observed two other groups of SPDBs: SPDBs-2 and SPDBs-3 (Fig.~\ref{fig_rspec}a). Some bursts in these SPDBs groups started even at frequencies of about 600 MHz (see e.g. the detailed spectra of SPDBs-2 in Fig.~\ref{r2}). The frequency drifts observed in groups of SPDBs-2 and SPDBs-3 were 50--180 MHz s$^{-1}$ and 56--100 MHz s$^{-1}$, respectively. The top part of  Table~\ref{table:1} provides summary information on the radio bursts observed in 600--2000 MHz range. 

At the frequency range of 2000--5000 MHz (Fig.~\ref{fig_rspec}b) we observed both continuum radio emission and bursts. Strong continuum emission was observed during the time intervals 06:57:40--06:58:20 UT and 06:59:20--07:00:40 UT. In addition to the continuum, two groups of narrow-band type III bursts were observed. The first group of narrow-band type III bursts, III-A, was observed at 06:57:48--06:57:56 UT at 4100--5000 MHz (Fig.~\ref{fig_rspec}b). This group consisted of three bursts separated in time by $\Delta$t$~\approx$\,4--5 s. All three bursts within this group were more intensive than the background continuum emission. The first and the last burst were observed approximately at 4500--5000 MHz and the middle one was observed through the largest frequency range 4100--4800 MHz. 
Another group of narrow-band type III bursts, III-B, was observed at 06:58:24--06:58:46 UT at frequencies of  4100--4800 MHz and was more complex than III-A. The III-B group consisted of several bursts where the last three bursts again seemed to be quasi-periodic in time with $\Delta$t$~\approx$\,4--5 s. The bottom part of   Table~\ref{table:1} provides summary information on the radio bursts observed in the 2000--5000 MHz range.

The time evolution of bursts described previously can also be seen in Fig.~\ref{fig_rspec}c. There, the cuts through the radio spectrum at three individual frequencies are plotted: 1190 MHz (brown), 4300 MHz (light blue), and 4680 MHz (orange). 

\begin{table*}
\caption{Summary of the observed groups of radio bursts.}
\label{table:1}
\centering     
\begin{tabular}{c c c c}
\hline\hline
Time range & Frequency range & Type & Characteristic\\
$[$UT$]$ & $[$MHz$]$ & & \\
\hline
    06:57:48--06:58:08 & 700--1300 & SPDBs-1 & freq. drift 54--90 MHz s$^{-1}$\\
    06:59:00--07:00:20 & 900--1400 & narrow-band III & seen over continuum\\
    07:01:08--07:01:47 & 600--1300 & SPDBs-2 & freq. drift 50--180 MHz s$^{-1}$\\
    07:02:05--07:02:50 & 600--1200 & SPDBs-3 & freq. drift 56--100 MHz s$^{-1}$\\

\hline
    06:57:40--06:58:20 & 2000--5000 & continuum & - \\
    06:57:48--06:57:56 & 4100--5000 & narrow-band III-A & 3 bursts seen over continuum\\
    06:58:24--06:58:46 & 4100--4800 & narrow-band III-B & several bursts seen over continuum\\
    06:59:20--07:00:40 & 2000--5000 & continuum & - \\
\hline
\end{tabular}
\end{table*}

\subsection{Hard X-rays}
\label{sec_hxr} 

Figure~\ref{fig_rspec}d shows the evolution of the HXR count rate by RHESSI during the time interval 06:57--07:03 UT when radio burst activity was detected. The HXR counts in 12--50 keV started to rise earlier than at 06:57 UT (not shown). In the  25--50 keV channel two main hard X-ray peaks were observed.
Generally, these peaks were well correlated in time with bands of continuum radio emission over the 2000--5000 MHz frequency range (Fig.~\ref{fig_rspec}b--d). The narrow-band type III-A (4100--5000 MHz) and also SPDBs-1 (700--1300 MHz) were observed during the maximum of the first 25--50 keV HXR peak.
Occurrence of the narrow-band type III-B (4100--4800 MHz) fell between the two main peaks of 25--50 keV HXR emission. However, it still correlates with small local peaks observed at 06:58:20--06:59:00 UT (Fig.~\ref{fig_rspec}c, d). Groups of SPDBs-2 and SPDBs-3 (07:01:08--07:01:47 UT and 07:02:05--07:02:50 UT) were observed already in the decline phase of HXR count rates. We note that the peak in count rates about 07:02 UT for both HXR energy intervals is artificial and occurred due to attenuator change. 

\subsection{Structure of the flaring active region: Magnetic field}
\label{sec_mag}

The photospheric magnetic structure of the flaring active region NOAA 12\,056, as observed by the SDO/HMI instrument, is shown in Fig.~\ref{fig_mag}a. The leading polarity was negative, consisting of three spots N1--3, several nearby plages and a large supergranule (LS). The following polarity has only one larger spot, P1, and several smaller ones together with more extended plages. Some of these plages are extended towards the west and south-west. The negative polarity spot, N1, was located north-west of P1 and N2 was located to the west of LS. The third sunspot of negative magnetic polarity, N3, was located about 100\arcsec~to the south of N1 (Fig.~\ref{fig_mag}a). From the east side, N3 was surrounded by an area of network magnetic field of positive polarity (Fig.~\ref{fig_mag}a) and on its west side, it was surrounded mostly by the magnetic field of negative polarity. EUV loops connecting N3 to a positive network magnetic field constitute a half dome magnetic structure (Fig.~\ref{fig_mag}b). A comparison of  Figs.~\ref{fig_mag}b and d shows  that during the flare there is  a faint ribbon surrounding the half dome structure in H$\alpha$.

%
\begin{figure}
\resizebox{\hsize}{!}{\includegraphics{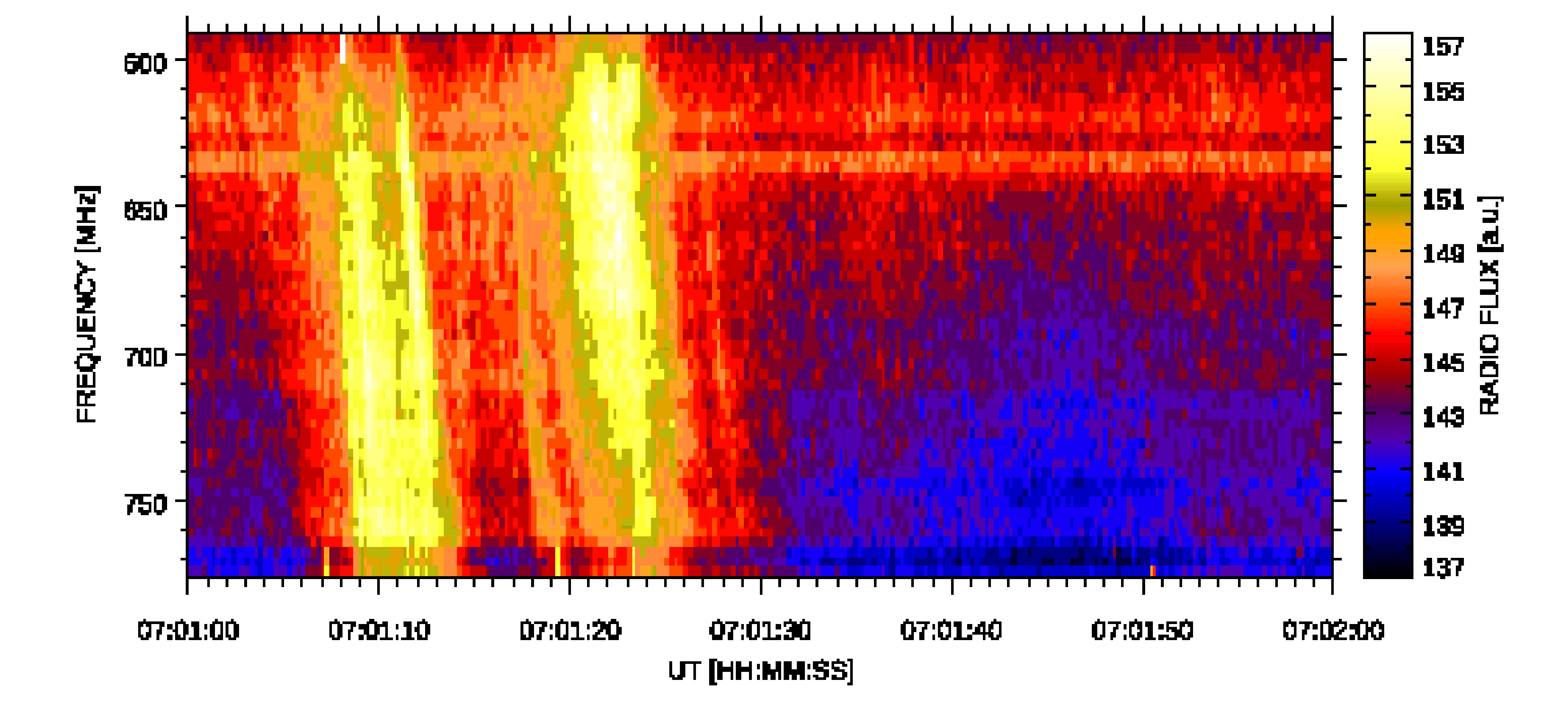}}
\resizebox{\hsize}{!}{\includegraphics{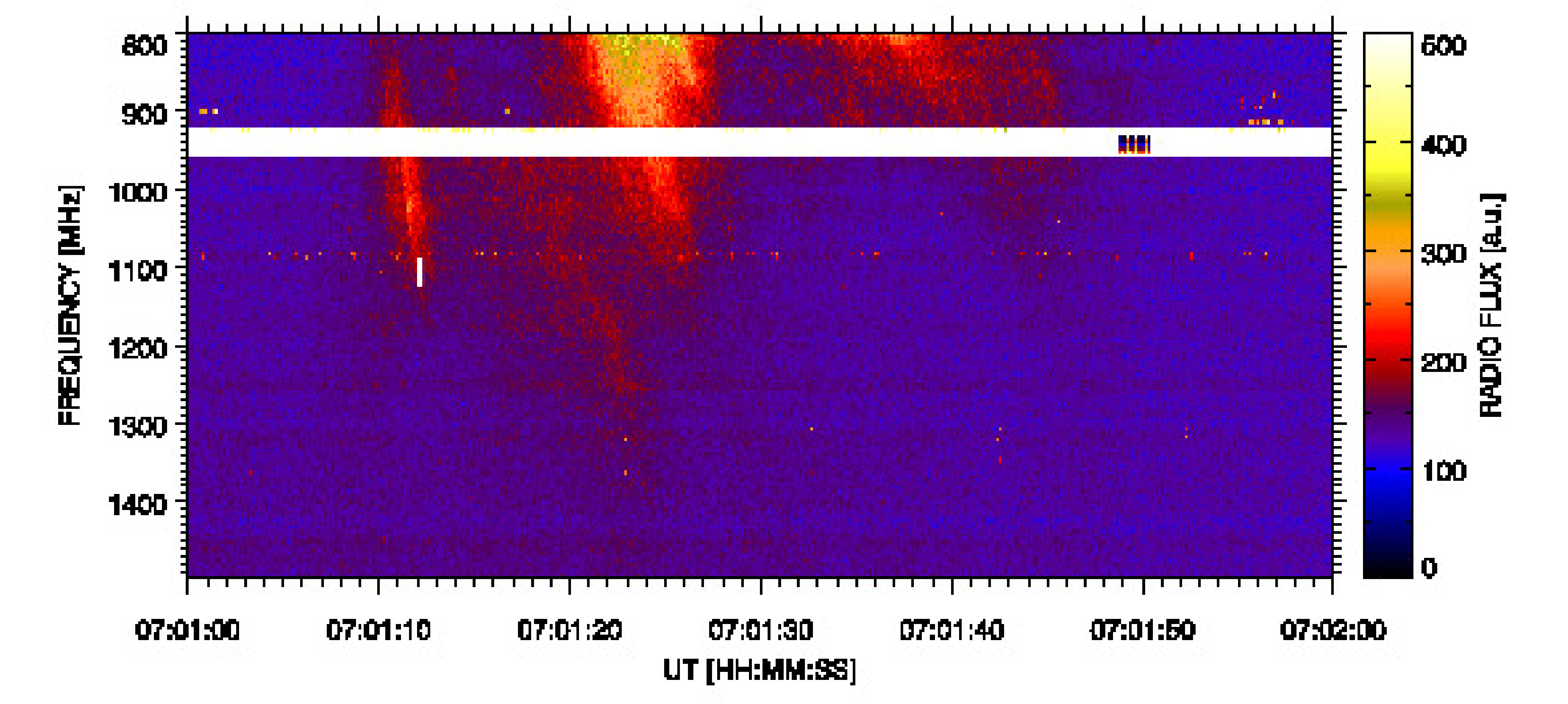}}
\caption{Detail of the radio spectrum showing SPDBs-2. The top panel is BLEN7M Callisto spectrum 
and the bottom panel shows spectrum from Ond\v{r}ejov (RT5). The panels have different scales 
in frequencies and radio fluxes. The short white vertical lines in the spectra indicate the times used in the
discussion for the velocity of beams generating SPDBs.}
 \label{r2}
\end{figure}
%
\begin{figure*}
\centering
\includegraphics[width=9.cm, clip, viewport= 2 45 378 348]{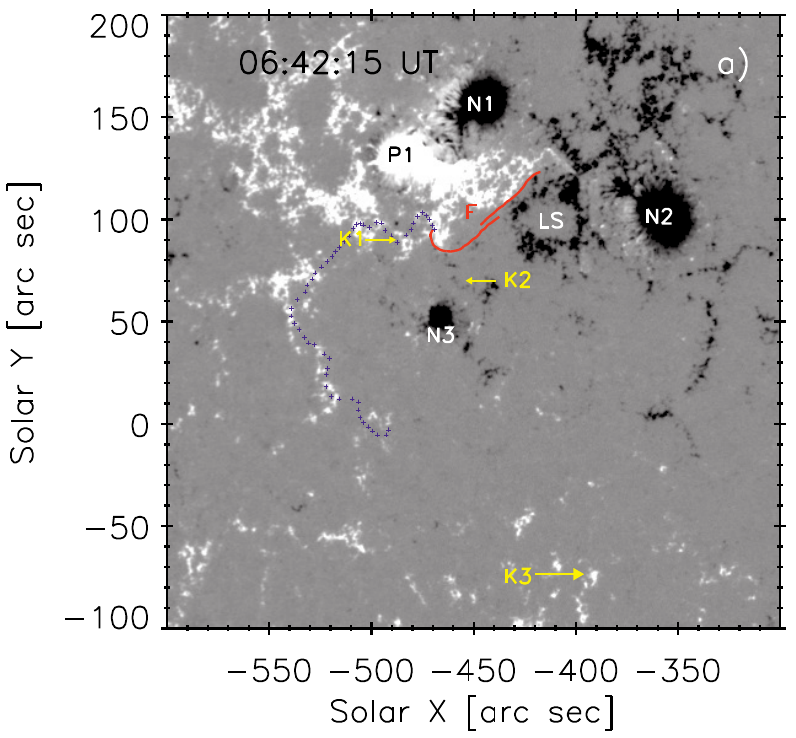}
\includegraphics[width=7.18cm, clip, viewport=78 45 378 348]{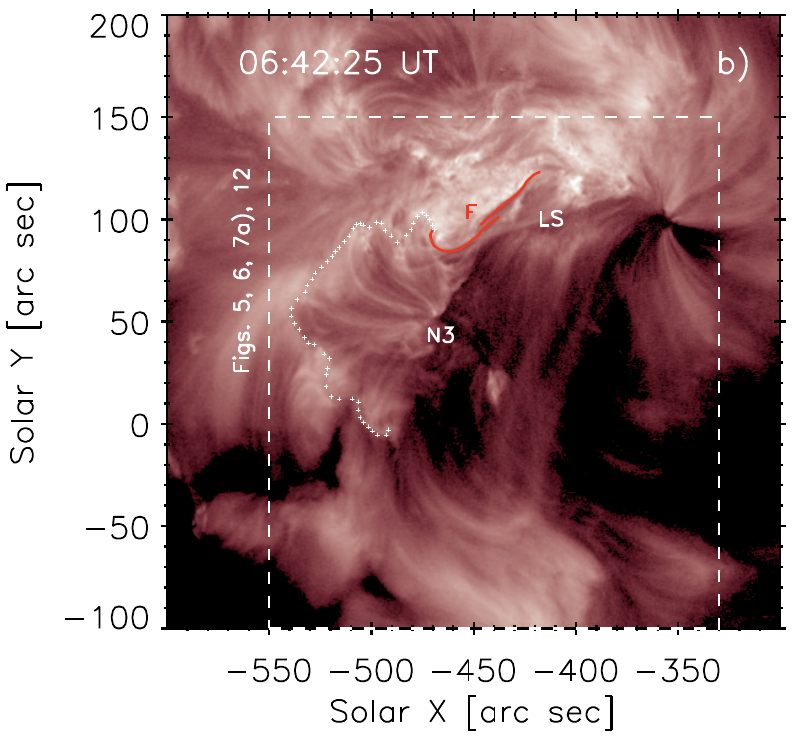}

\includegraphics[width=9.cm, clip, viewport= 2 0 378 348]{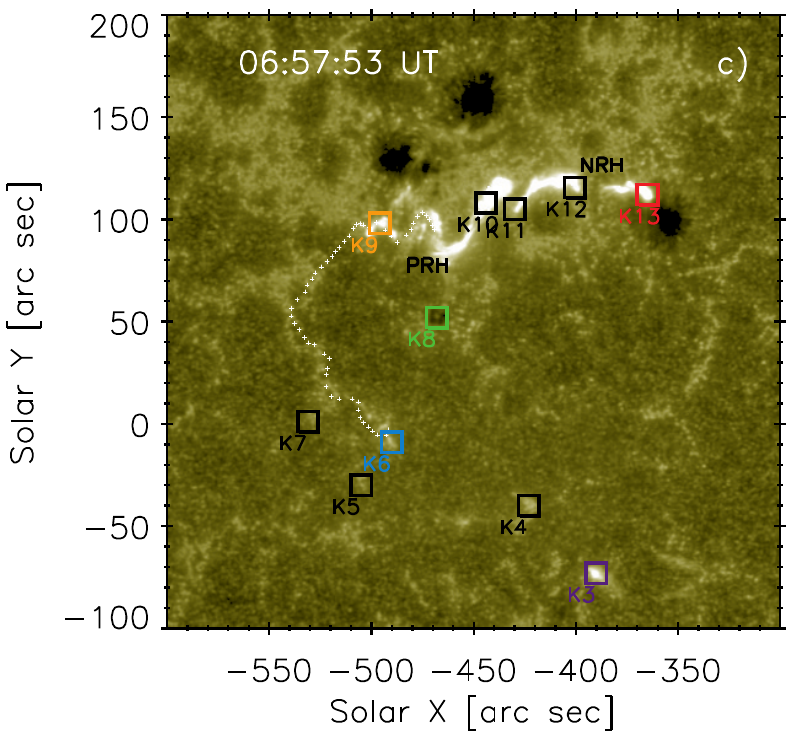}
\includegraphics[width=7.18cm, clip, viewport=78 0 378 348]{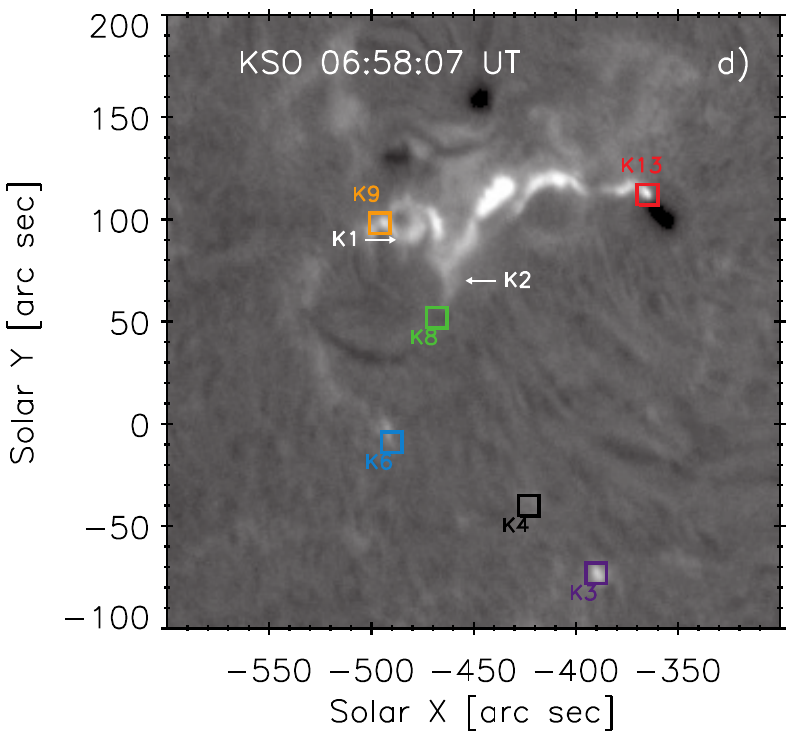}

\caption{Overview of active region and flare kernels.
(a) SDO/HMI line-of-sight magnetogram ($\pm$ 300\,G): P1, N1, N2, N3 - sunspots; LS - large supergranule; F  - filament (red lines); K1, K2, and K3  - flare kernels (yellow arrows); and  border of the half dome structure (tiny dark blue crosses).
(b) SDO/AIA image in 211 \AA~showing the loops of the half dome and filament F (red lines). The border of the half dome is shown by white crosses (also in (c)). The dashed rectangle is the field of view for Figs.~\ref{fig_flare_1},~\ref{fig_reco}, ~\ref{fig_hl}a, and~\ref{fig_schema}.
(c) SDO/AIA 1600  \AA~image showing ribbons and their positive and negative polarity ribbon hooks (PRH and NRH, respectively). The coloured and numbered squares show the locations of particular flare kernels.
(d) KSO H$\alpha$ image showing the positions of the flare kernels at the beginning of the flare (K1, K2 - arrows), and those resembling the model of 3D magnetic reconnection (K3, K4, K6, K8, K9, K13).}
    \label{fig_mag}
\end{figure*}

\subsection{Evolution of the flaring active region in EUV and H$\alpha$}
\label{sec_euv}

We now  describe the evolution of the flare as observed in EUV and H$\alpha$. In doing so, we give particular emphasis to several areas of bright footpoint emission, labelled as kernels K1--K13. The individual kernels brighten at different times and their locations are denoted in Fig. \ref{fig_mag}c. Their evolution in relation to the main flaring processes is described in the remainder of this section, while their correlations to the radio emission (or lack thereof) are studied  in Sect.~\ref{sec_uv_lc}.

\subsubsection{Start of the flare}
\label{sec_flare_start}

Along the polarity inversion line within the active region NOAA 12\,056, a small active region filament F was observed. Its location is shown in Fig.~\ref{fig_mag}a and b by a red line,
with the filament F being most readily visible in SDO/AIA 304 \AA~(Fig.~\ref{fig_flare_1}c.) The activation of  filament F started at about 06:45 UT (see the movie associated with Fig.~\ref{fig_flare_1}). At 06:47--06:51 UT, we observed the formation of a bright sigmoid (S) in the hot AIA passband overlying  filament F (Fig.~\ref{fig_flare_1}a). The first bright flare kernel (K1) (Fig.~\ref{fig_flare_1}a--c) appeared at about 06:49 UT in the extended positive magnetic field (Fig.~\ref{fig_mag}a) close to the eastern footpoint of S.  The first flare kernel (K1) was located in the northern part of the half dome (Fig.~\ref{fig_flare_1}a). At this time, we observed two loops, L1 and L2, overarching  filament F in the SDO/AIA 171 \AA~filter (Fig.~\ref{fig_flare_1}b). 
The second flare kernel (K2) appeared near the `elbow' of S (Fig.~\ref{fig_flare_1}a), located in a weak magnetic field of negative polarity north of N3 (Fig.~\ref{fig_mag}a). In the hot 131 \AA~EUV channel (Fig.~\ref{fig_flare_1}a) we observed tiny loops connecting kernels K1 and K2. At about 06:50 UT, the negative footpoints of these loops connecting K1 and K2 started to slip from K2 towards N3.

\subsubsection{Evolution of the flare core}
\label{sec_flare_core}

\begin{figure*}[ht!]
\centering
\includegraphics[width=7.0cm, clip, viewport=2  48 340 348]{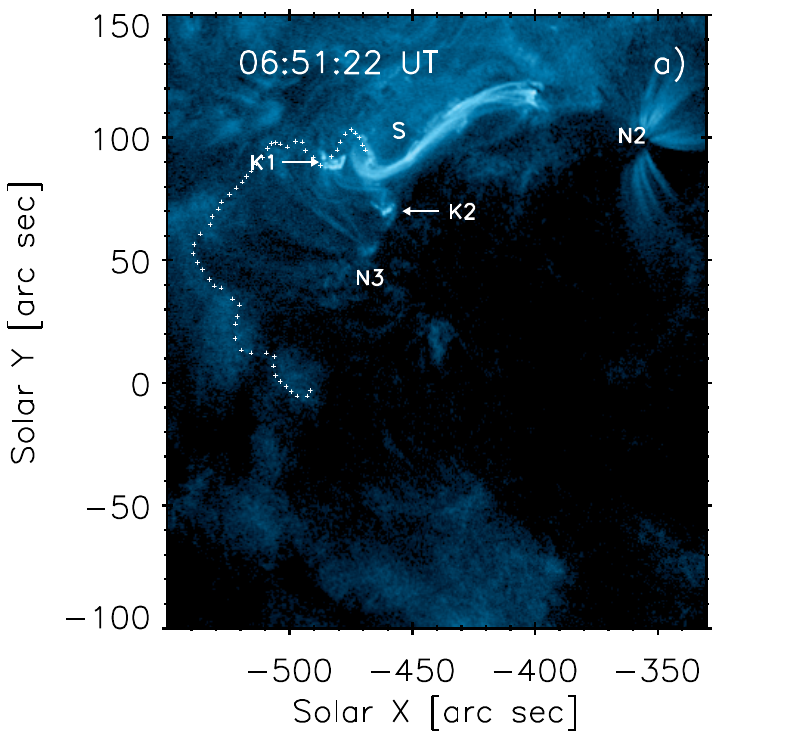}
\includegraphics[width=5.425cm, clip, viewport=78 48 340 348]{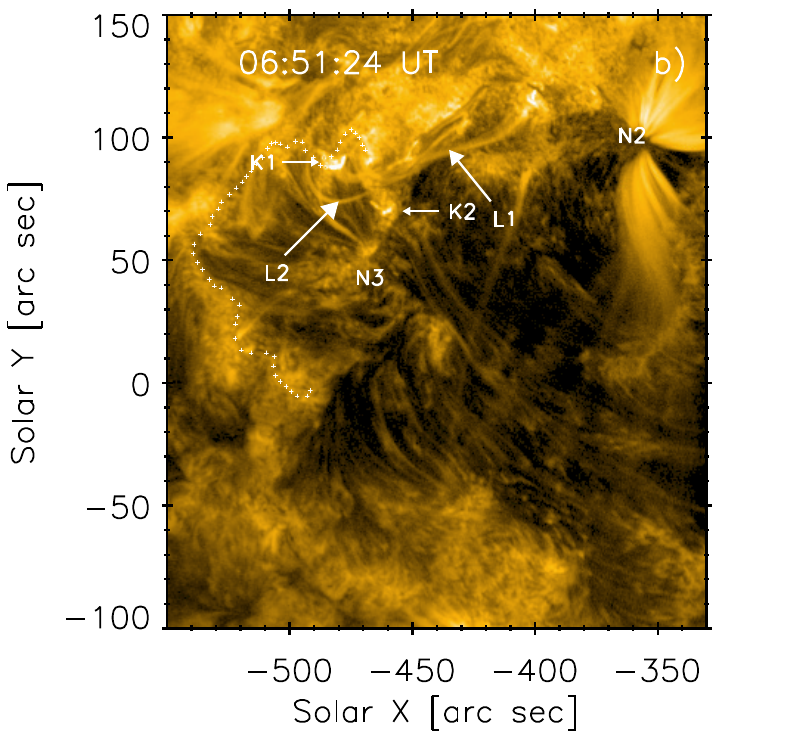}
\includegraphics[width=5.425cm, clip, viewport=78 48 340 348]{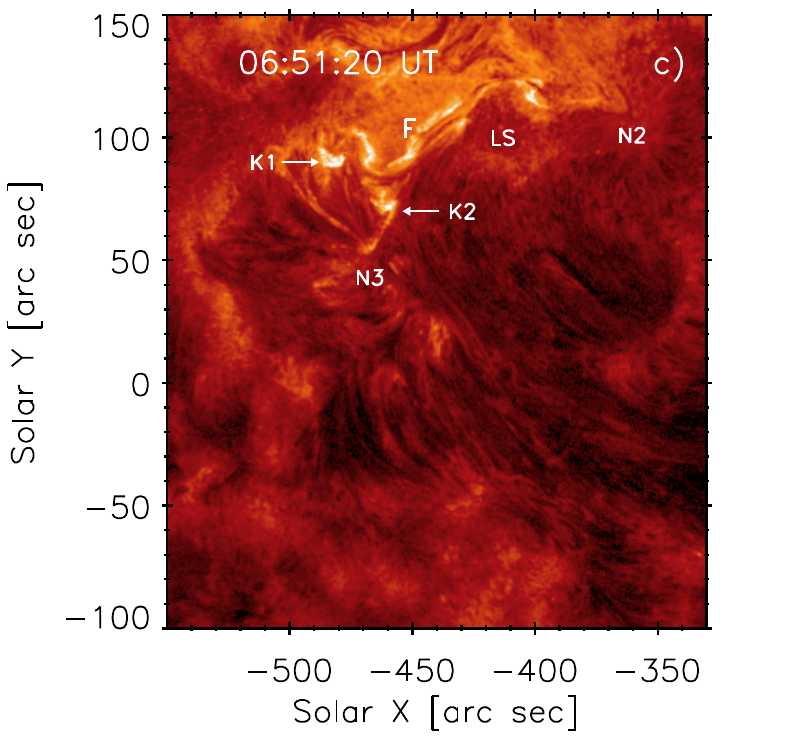}

\centering
\includegraphics[width=7.0cm,clip, viewport=2  48 340 348]{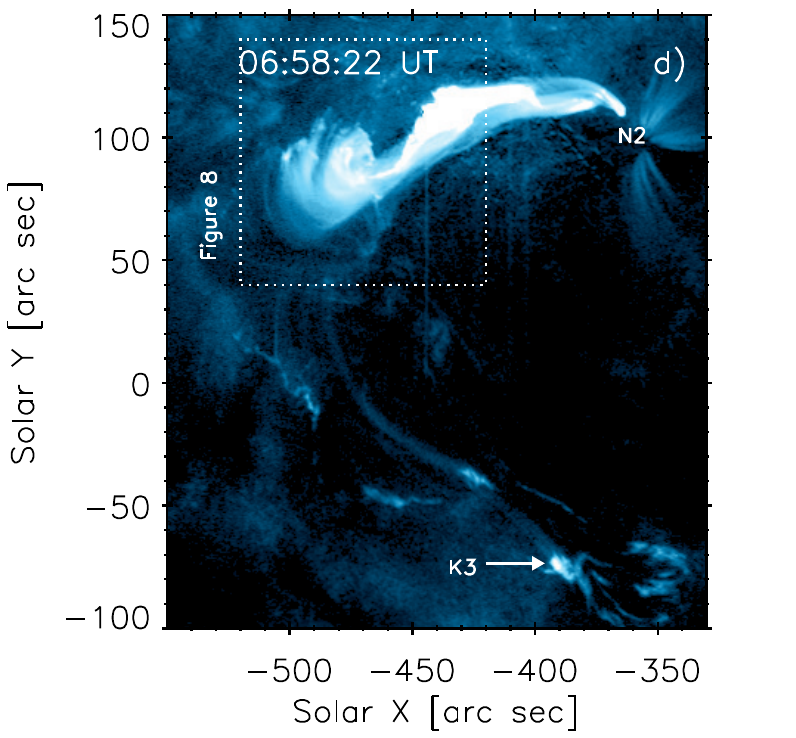}
\includegraphics[width=5.425cm,clip, viewport=78 48 340 348]{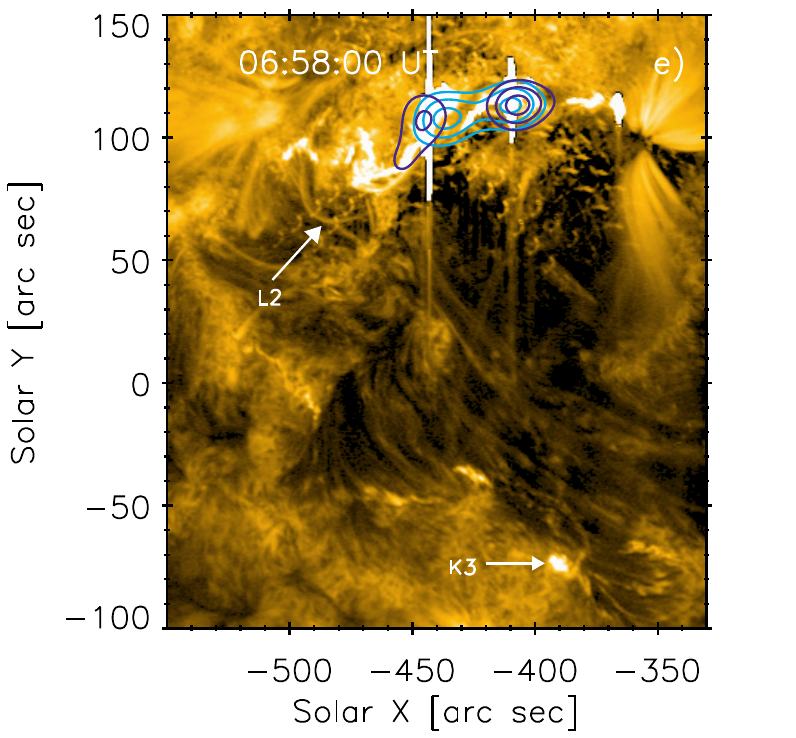}
\includegraphics[width=5.425cm,clip, viewport=78 48 340 348]{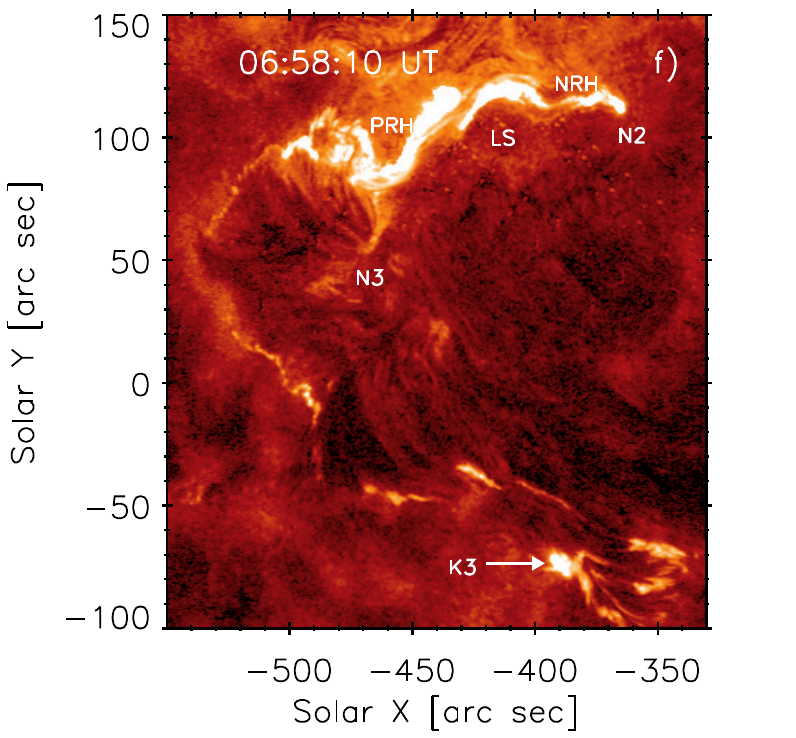}

\centering
\includegraphics[width=7.0cm,clip, viewport=2  0 340 348]{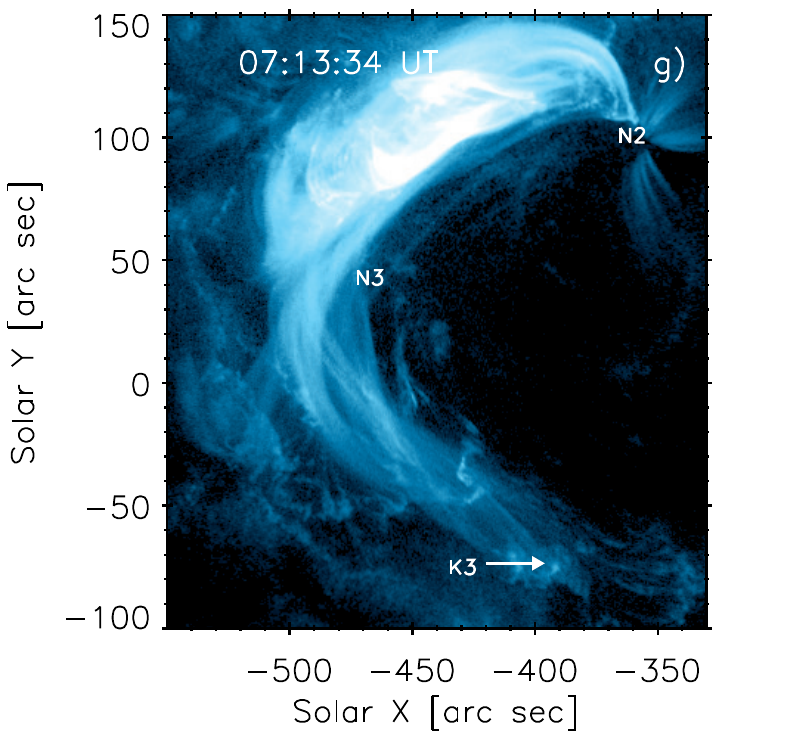} 
\includegraphics[width=5.425cm,clip, viewport=78 0 340 348]{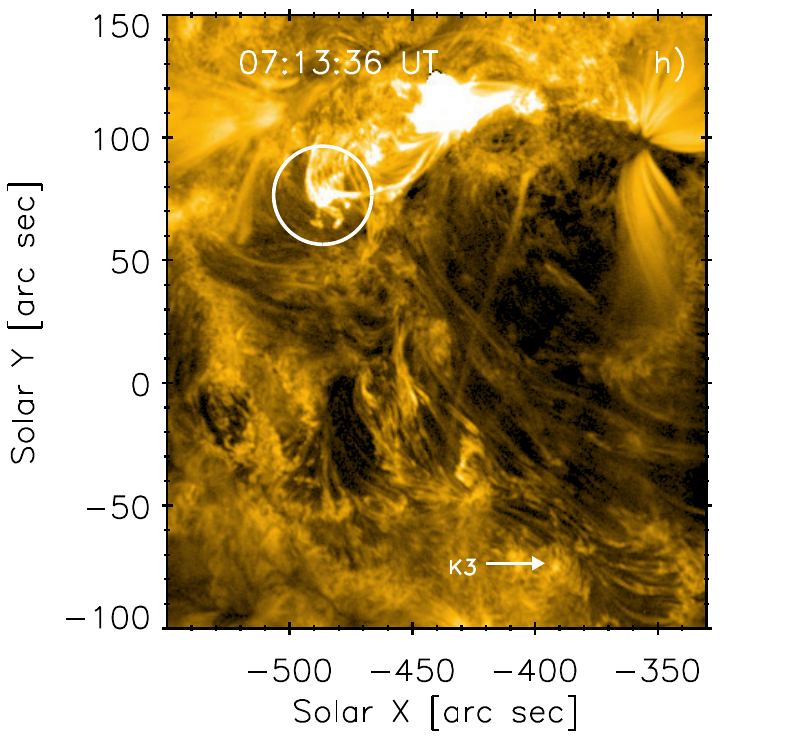} 
\includegraphics[width=5.425cm,clip, viewport=78 0 340 348]{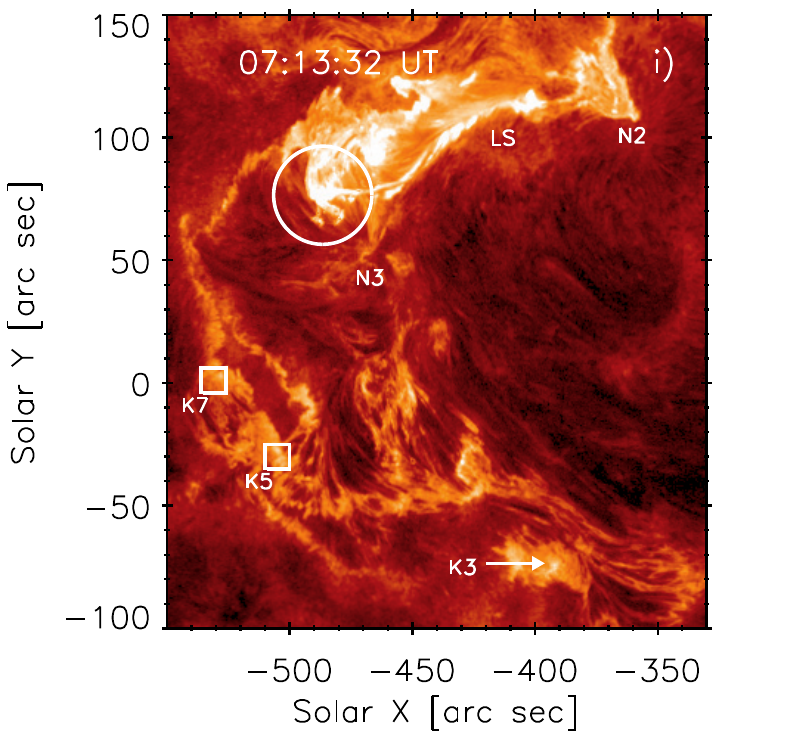} 
\caption{SDO/AIA images showing the evolution of the C8.7 flare in EUV: 131 \AA~filter (a, d, g), 171 \AA~(b, e, h), and 304 \AA~(c, f, i). N2 and N3 give the  positions of negative sunspots; LS is the  large supergranule; S is the bright sigmoid formed over the filament F; and PRH and NRH in (f) are ribbon hooks. The tiny white crosses show the border of the half dome (a and b), the arrows point to the flare kernels K1, K2 in (a--c), and K3 in (d--i), and to loops L1 and L2 in (b and e); and the white circles in (h and i) show the position of the swirl. The dotted rectangle in (d) shows the field of view in Fig.~\ref{fig_swirl}. The contours in (e) show RHESSI HXR sources at 06:58:34--06:59:28 UT: light blue 12--25 keV, dark blue 25--50 keV. The temporal evolution of the flare (06:45--07:45 UT) in SDO/AIA filters 131 \AA, 171 \AA, 304 \AA,~and 1600 \AA~is available as an online movie (flare\_131\_171\_304\_1600.mpg).}
    \label{fig_flare_1}
\end{figure*}

\begin{figure*}
\includegraphics[width=7.cm,clip, viewport=2 0 340 348]{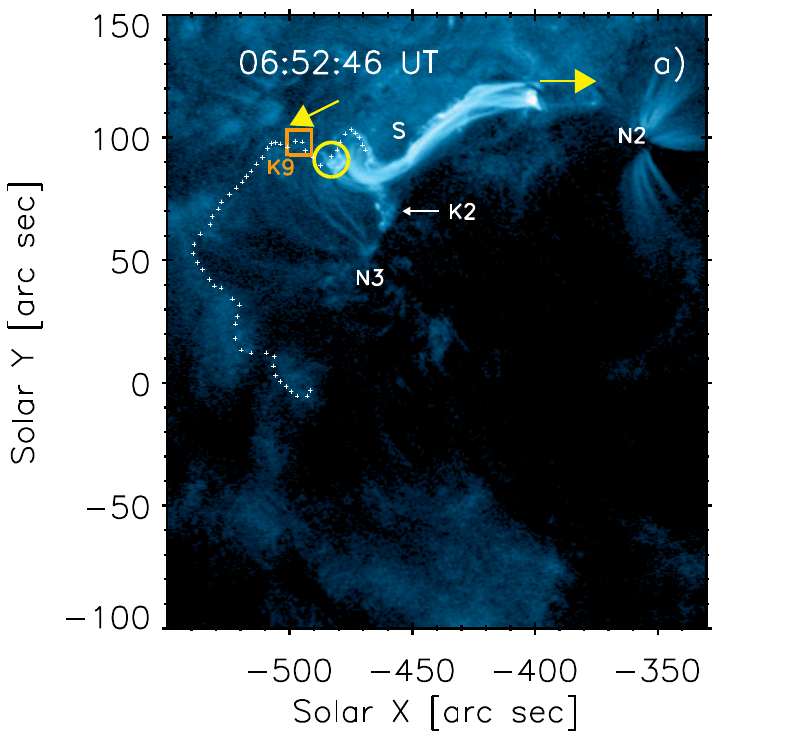}
\includegraphics[width=5.425cm,clip, viewport=78 0 340 348]{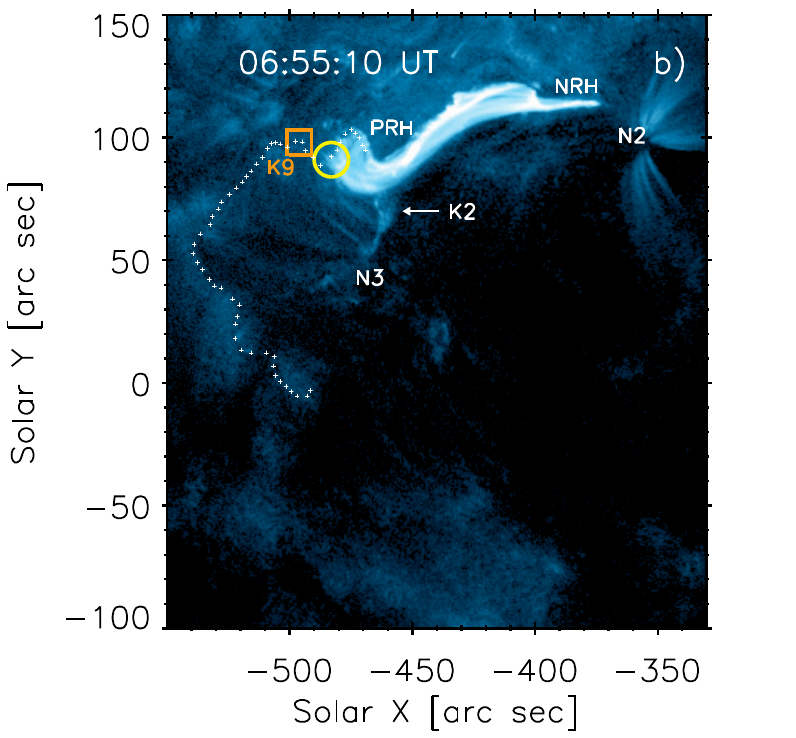}
\includegraphics[width=5.425cm,clip, viewport=78 0 340 348]{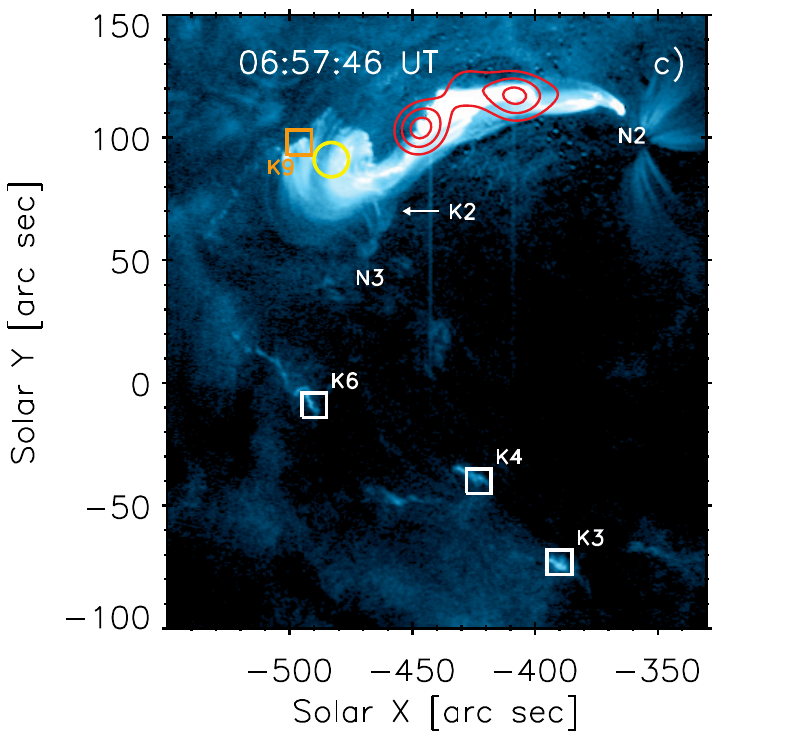}
\caption{Interaction of S with the half dome. The tiny white crosses show the border of the half dome in (a) and (b). 
In each panel N2 and N3 give the positions of the negative sunspots, K1 is located at the centre of the yellow circle, the white arrow points to K2, and the orange square at the border of the half dome locates K9. The yellow arrows in (a) denote the direction of the evolution of the hooks PRH and NRH (shown in (b)); the white squares show K3, K4, and K6. The red contours in (c) show the positions of HXR (25--50 keV) at 06:57:40--06:58:33 UT with 50, 70, and 90\% of the maximum.}
   \label{fig_reco}
\end{figure*}

Simultaneously, at about 06:50 UT, the S started to grow. Its growth was associated with the appearance of a pair of J-shaped ribbons (Fig.~\ref{fig_flare_1}e, f). The growing S first interacted with overlying loops L1 and L2, and later with the half dome. Figure~\ref{fig_reco} shows more details of this interaction process adding three time instants between those shown in Fig.~\ref{fig_flare_1}a and d. Meanwhile, as the  S was growing, its positive ribbon hook (PRH) was widening and the footpoints of the associated flare loops crossed the location of K1 (Fig.~\ref{fig_reco}, yellow circle). They continued slipping further towards the position of the orange square denoting kernel K9 (Fig.~\ref{fig_reco}a--c) following the border of the half dome from its outer side. The conjugate negative ribbon hook (NRH)
was drifting towards N2. The directions of evolution of the  PRH and NRH are shown by yellow arrows above S in Fig.~\ref{fig_reco}a. Approximately at 06:57 UT, NRH reached the penumbra-umbra border of N2 sunspot (Fig.~\ref{fig_reco}c) and the loops of PRH reached  kernel K9 (Fig. \ref{fig_reco}). 
After this time, several new bright kernels (K3, K4, and K6), appeared to the south of N3 (Fig.~\ref{fig_reco}c) and later also new large-scale flare loops were observed there (see Sect. \ref{sec_K3}). 
Meanwhile, a pair of RHESSI HXR sources at 25--50 keV appeared along the straight parts of the  J-shaped ribbons (dark blue contours in Fig.~\ref{fig_flare_1}e). Figure~\ref{fig_flare_1}d--e shows that these bright straight parts of the ribbons were joined by hot flare loops of the central part of S, later seen to be cooling at 304 \AA~(Fig.~\ref{fig_flare_1}i). 

%

%
\begin{figure*}
\includegraphics[width=7.cm, clip, viewport=2 0 340 348]{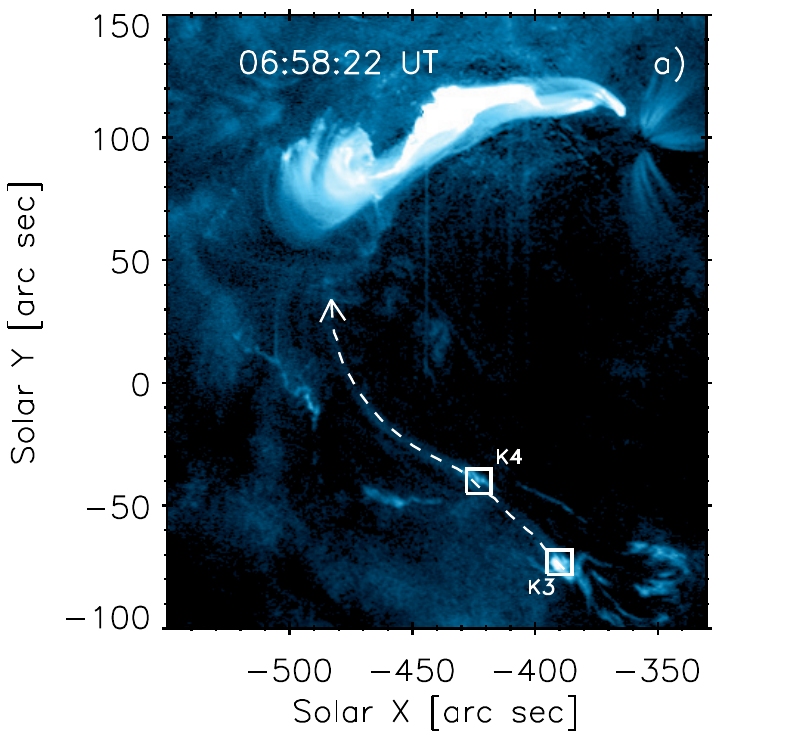}
\includegraphics[width=11.5cm, clip, viewport=2 4 508 280]{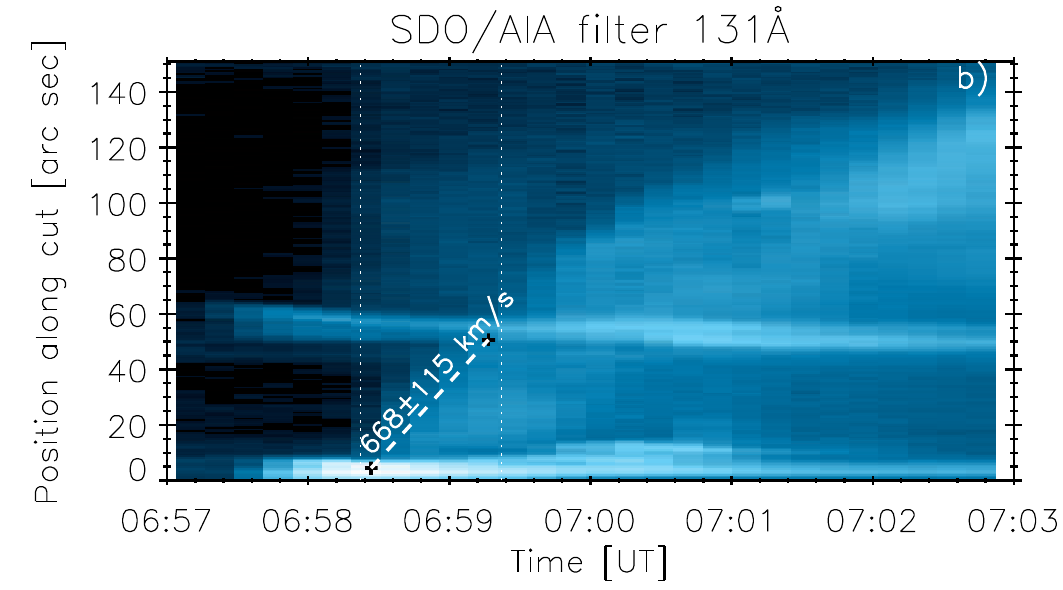}
\caption{Evaporation of hot plasma. (a) Path of the cut starting from K3 and ending by an arrowhead. 
(b) Time-distance plot showing evaporation of hot plasma filling the new flare loop from K3. 
The velocity of the evaporation is indicated.}
    \label{fig_hl}
\end{figure*}

\subsubsection{Outlying kernels and large-scale flare loops}
\label{sec_K3}

During the interaction of S with the half dome (Sect.~\ref{sec_flare_core}), several new kernels appeared further away from the flaring active region, mostly to the south of N3. These included a bright kernel K3 (Fig.~\ref{fig_mag}c), and a few fainter ones, including K4 and K6 (Fig. \ref{fig_reco}c). Radio emission in 600--5000 MHz was observed almost simultaneously with the appearance of those kernels. From 07:00 UT other flare kernels,   K5 and K7, were also appearing to the south and  south-east from N3 (Fig.~\ref{fig_flare_1}i). 

At 06:58 UT,  kernel K3 was the brightest and southernmost  of these (Fig.~\ref{fig_flare_1}d--f). It was located in positive polarity far to the south of the main flare, more than 100\arcsec~south from N3 (Fig.~\ref{fig_mag}a). K3 was visible in emission also in UV (Figs.~\ref{fig_mag}c) and in H$\alpha$ (Fig.~\ref{fig_mag}d). We note that details on the H$\alpha$ evolution of this flare can be found in \citet{Tschernitz2018}. 

Figure~\ref{fig_flare_1}d also shows that slightly after the appearance of kernel K3 a new large-scale flare loop emanating from it started to be seen. We attempted to estimate the  velocity of the plasma evaporation filling this large-scale flare loop from the side of K3. We traced a part of it, as shown by the dashed line in Fig.~\ref{fig_hl}a. A time-distance technique was used to measure the image-plane projected evolution of hot plasma along this trace (Fig.~\ref{fig_hl}b). As the path starting from K3 passed over  kernel K4 as well (Figs.~\ref{fig_hl}a, \ref{fig_flare_1}d), for estimation of evaporation velocity we used only the part of the loop between the two kernels (Fig.~\ref{fig_hl}b). This way, we excluded emission from other flare loops that might originate from K4. The measured projected velocity 668 $\pm$~115 km.s$^{-1}$ is among the highest evaporation velocities reported to date \citep[see e.g.][]{Antonucci1990,Doschek2013,Young2013,Polito2016,Dudik2016,Lee2017}. This could indicate a significant energy deposit into the bright kernel K3, although we did not observe any HXR source there.

Later we observed a bundle of hot loops connecting the area around K3 with the N2 spot (Fig.~\ref{fig_flare_1}g). In addition to this bundle of hot loops, there are also two other systems of large-scale flare loops starting from N2. One connects N2 with a system of new extended flare ribbons located approximately at coordinates solar-X $\sim$ [$-550$\arcsec -- $-490$\arcsec], solar-Y $\sim$ [$-50$\arcsec -- 0\arcsec], including kernels K5 and K7 located at these ribbons (Fig.~\ref{fig_flare_1}g--i). Another large-scale system of flare loops connects N2 with the north-eastern part of the half dome border. It is composed of flare loops overlying the core flare arcade (Fig.~\ref{fig_flare_1}g). The presence of all these large-scale flare loops indicates significant expansion of the connectivity of the flare from its original location at the S.

\subsubsection{Extreme ultraviolet swirl}
\label{sec_swirl}

\begin{figure*}
\centering
\includegraphics[width=5.52cm, clip, viewport=2 45 378 348]{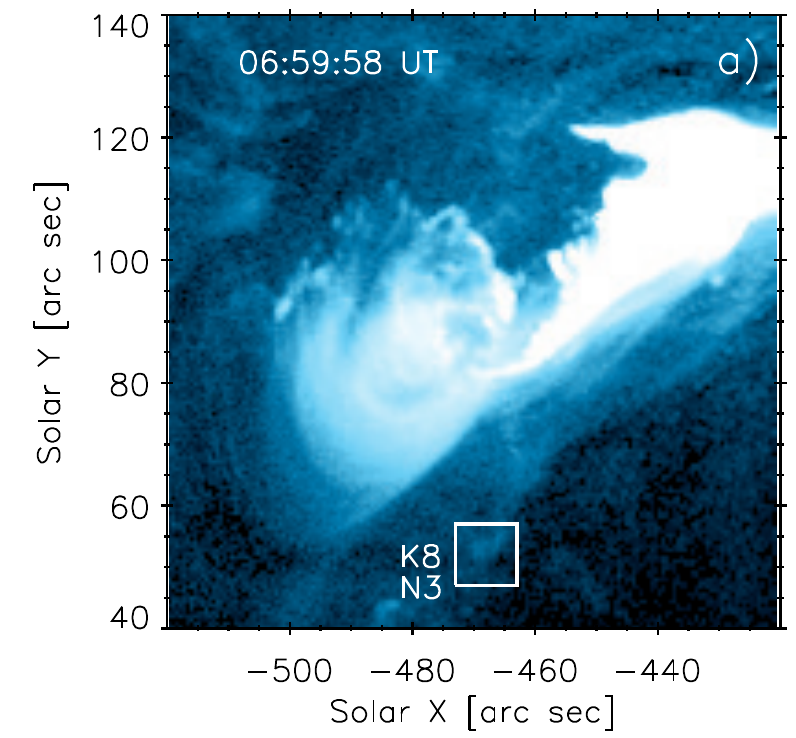} 
\includegraphics[width=4.4cm, clip, viewport=78 45 378 348]{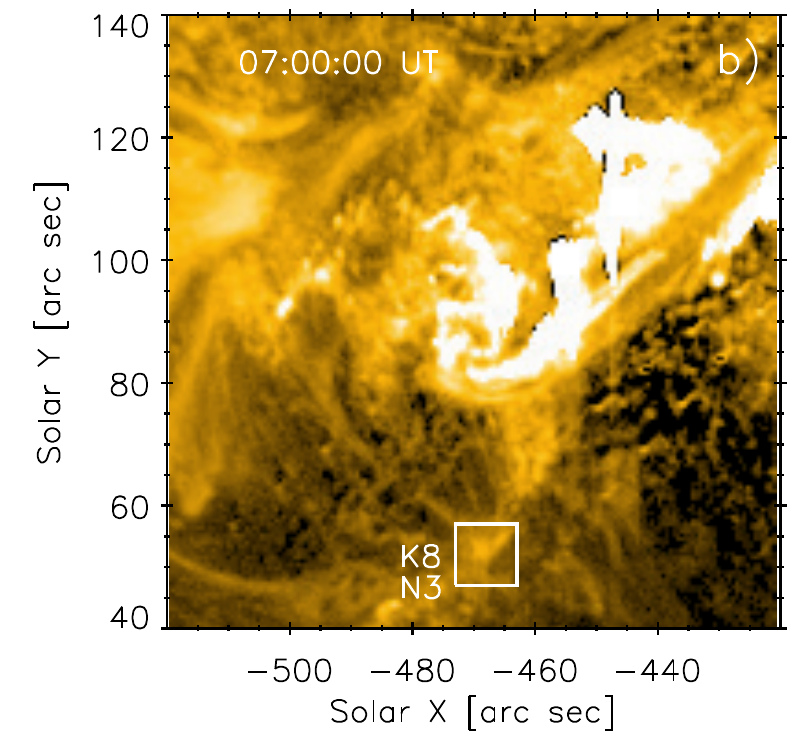} 
\includegraphics[width=4.4cm, clip, viewport=78 45 378 348]{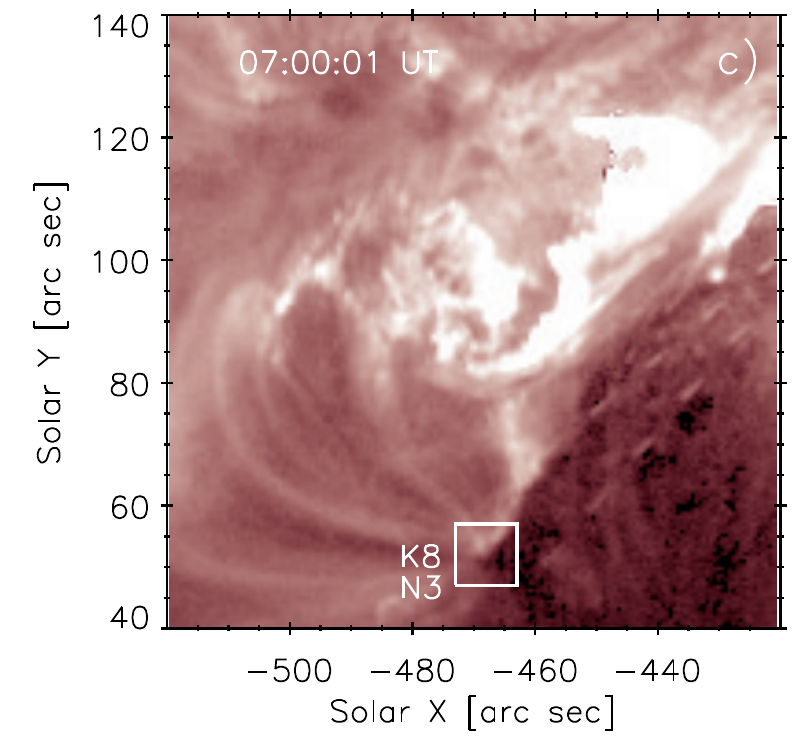} 

\includegraphics[width=5.52cm, clip, viewport=2 45 378 348]{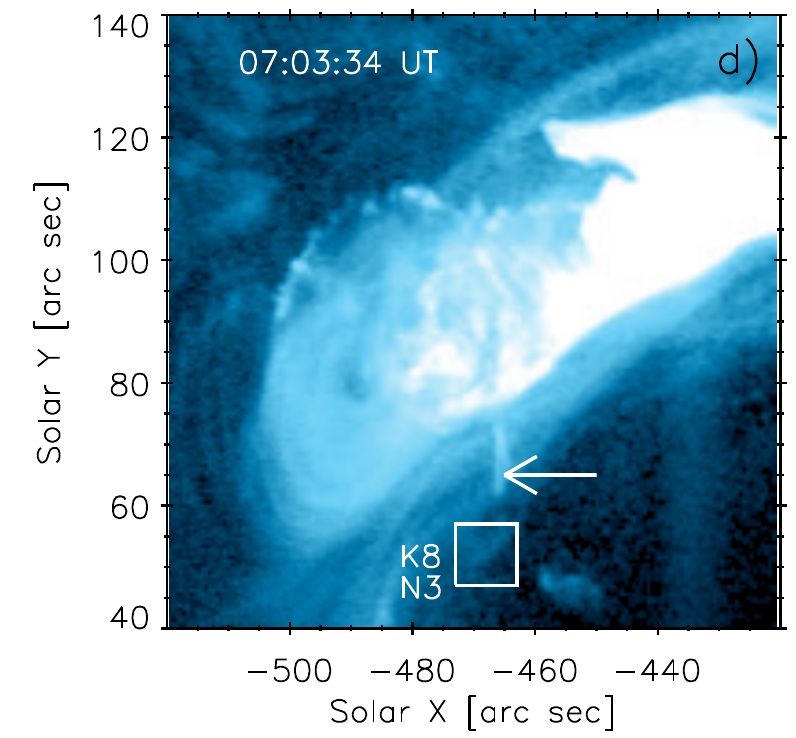} 
\includegraphics[width=4.4cm, clip, viewport=78 45 378 348]{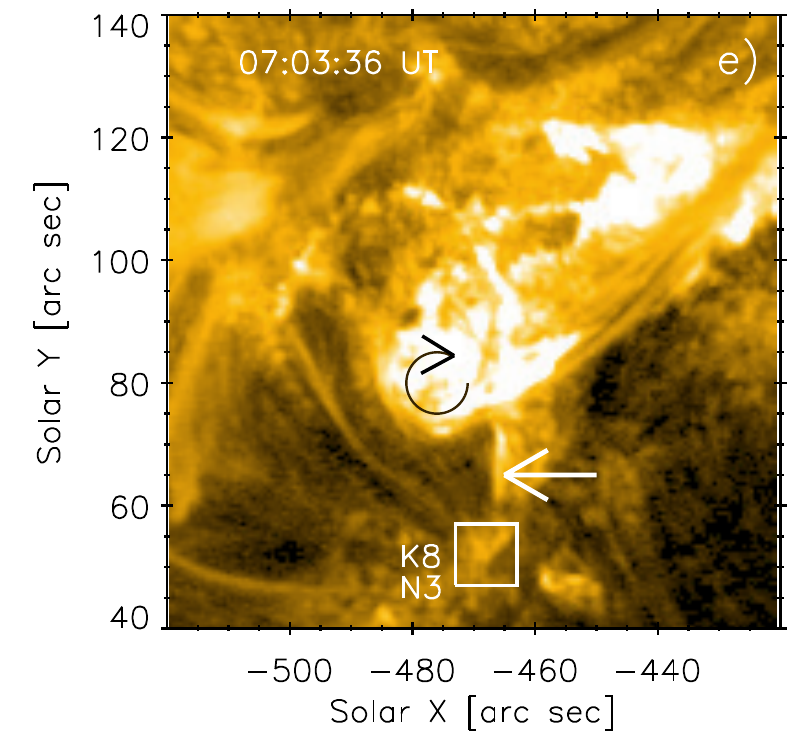} 
\includegraphics[width=4.4cm, clip, viewport=78 45 378 348]{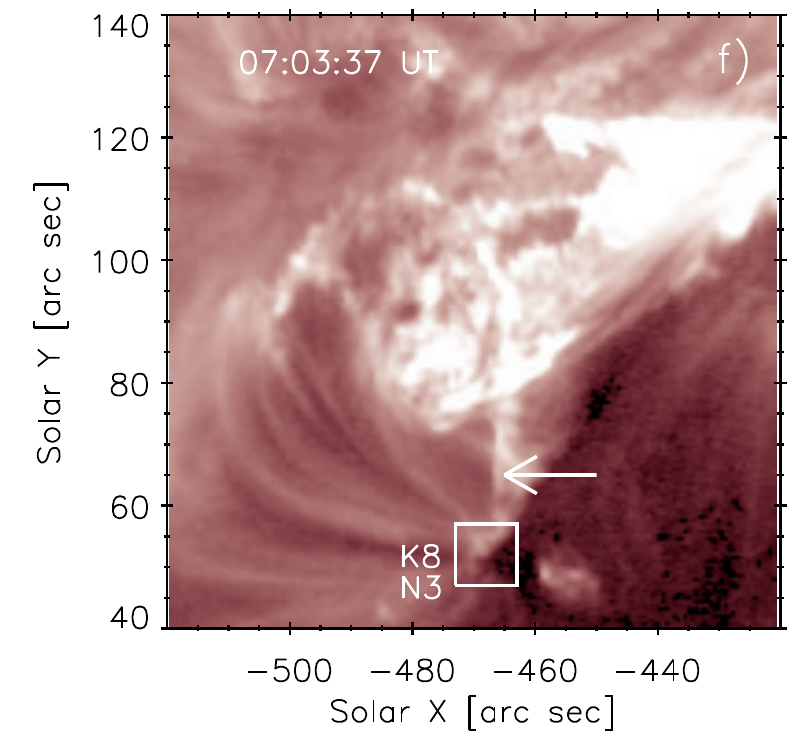} 

\includegraphics[width=5.52cm, clip, viewport=2 45 378 348]{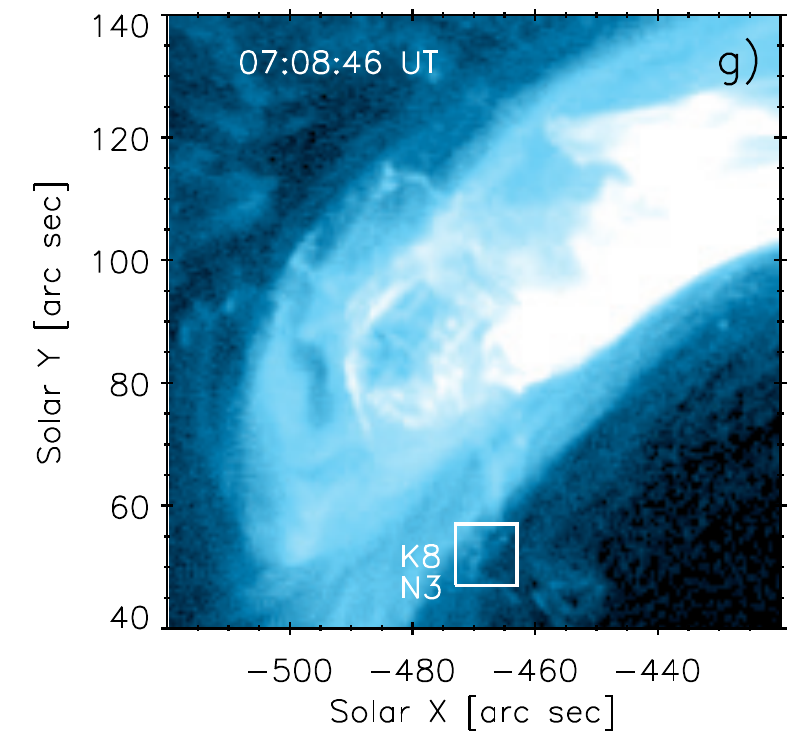} 
\includegraphics[width=4.4cm, clip, viewport=78 45 378 348]{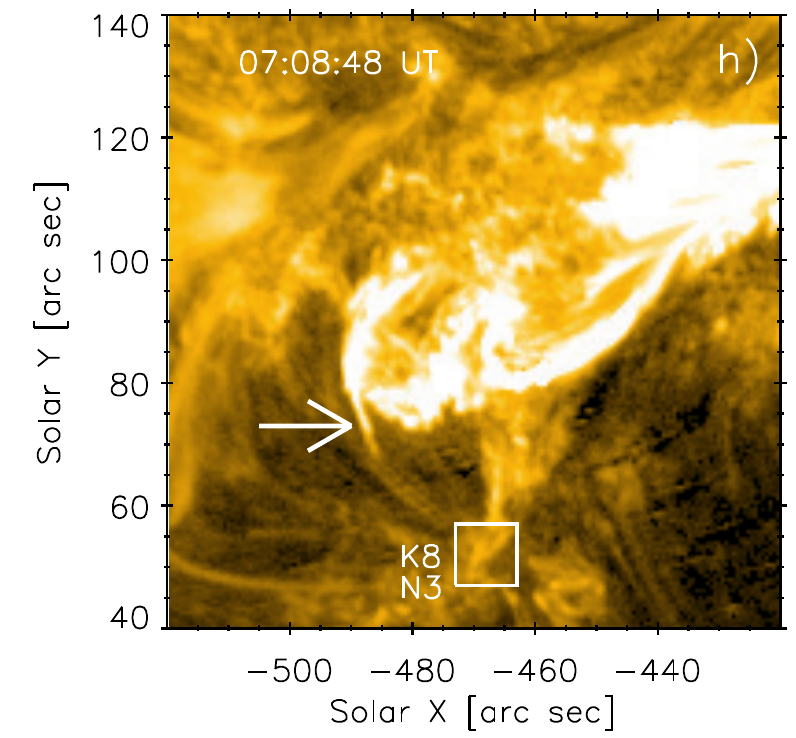} 
\includegraphics[width=4.4cm, clip, viewport=78 45 378 348]{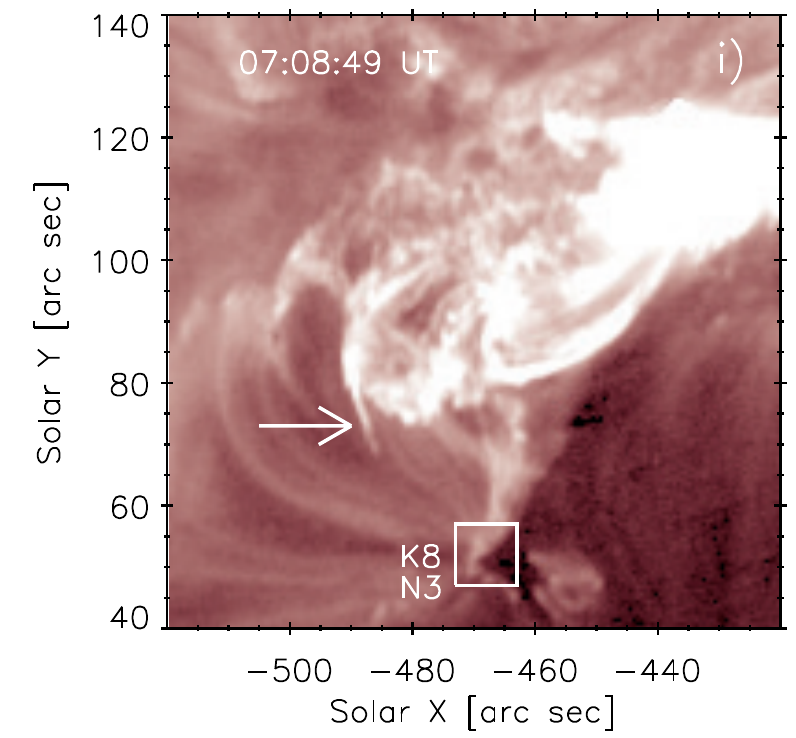} 

\includegraphics[width=5.52cm, clip, viewport=2 45 378 348]{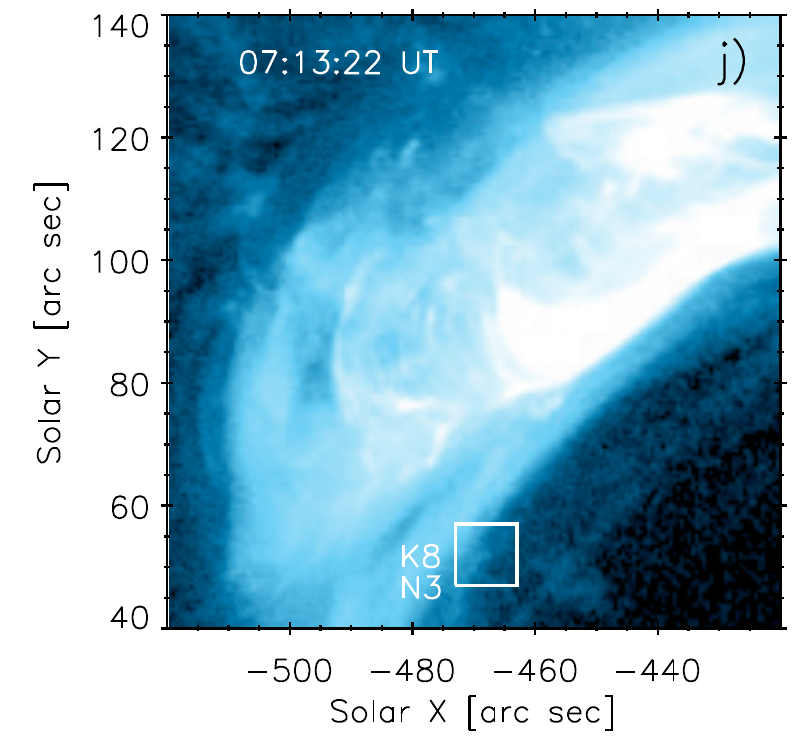} 
\includegraphics[width=4.4cm, clip, viewport=78 45 378 348]{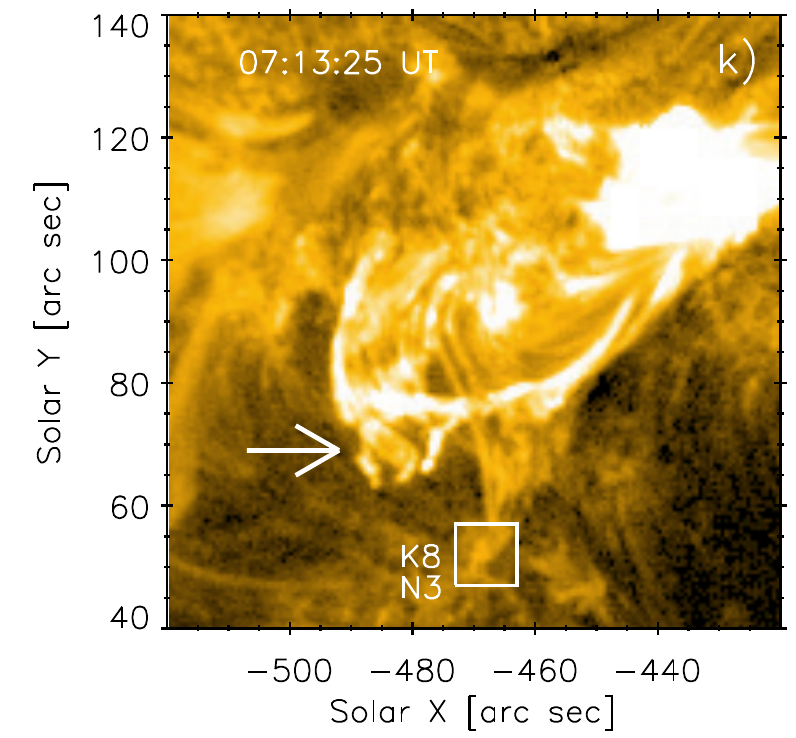} 
\includegraphics[width=4.4cm, clip, viewport=78 45 378 348]{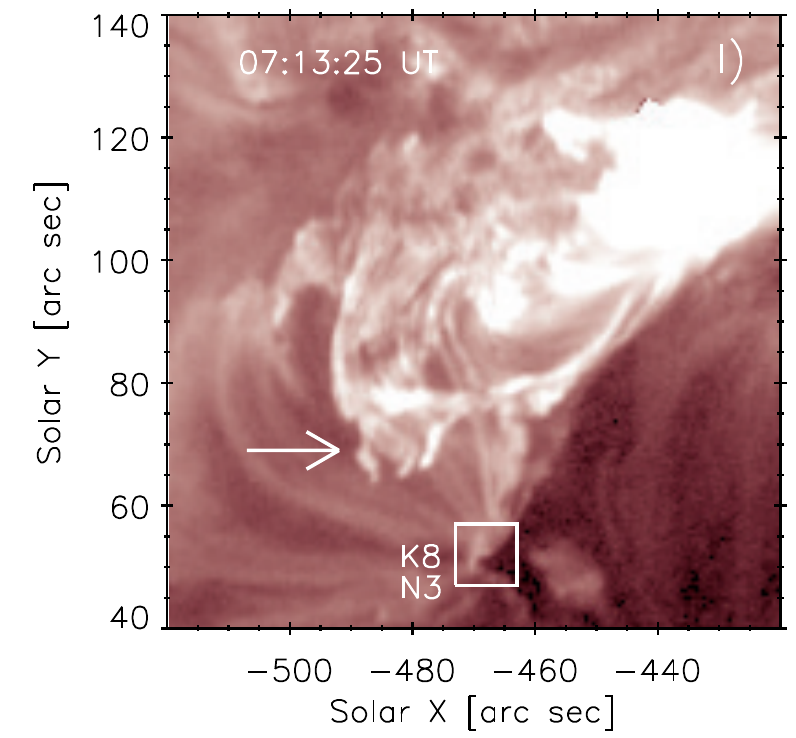} 

\includegraphics[width=5.52cm, clip, viewport=2 0 378 348]{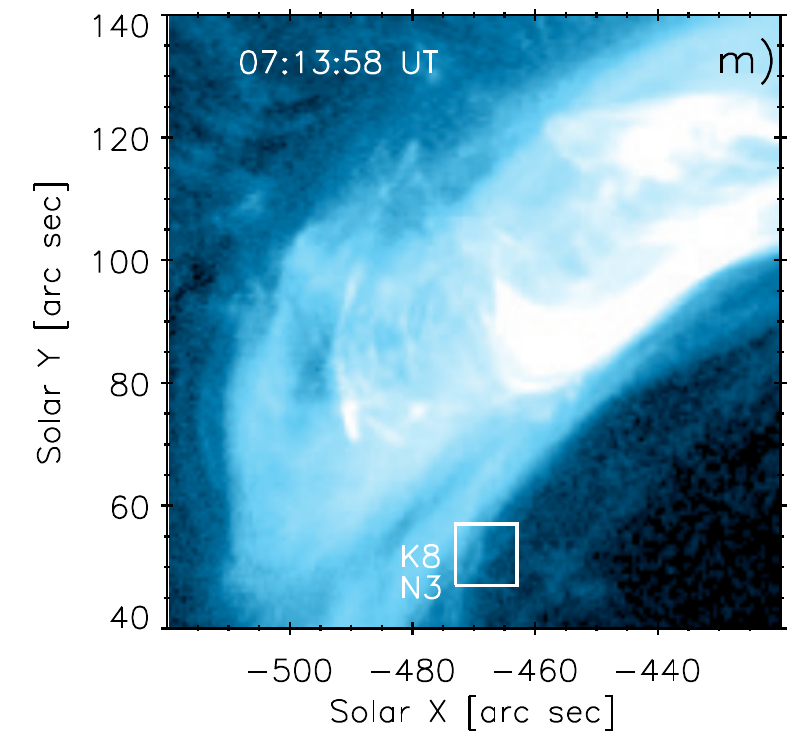} 
\includegraphics[width=4.4cm, clip, viewport=78 0 378 348]{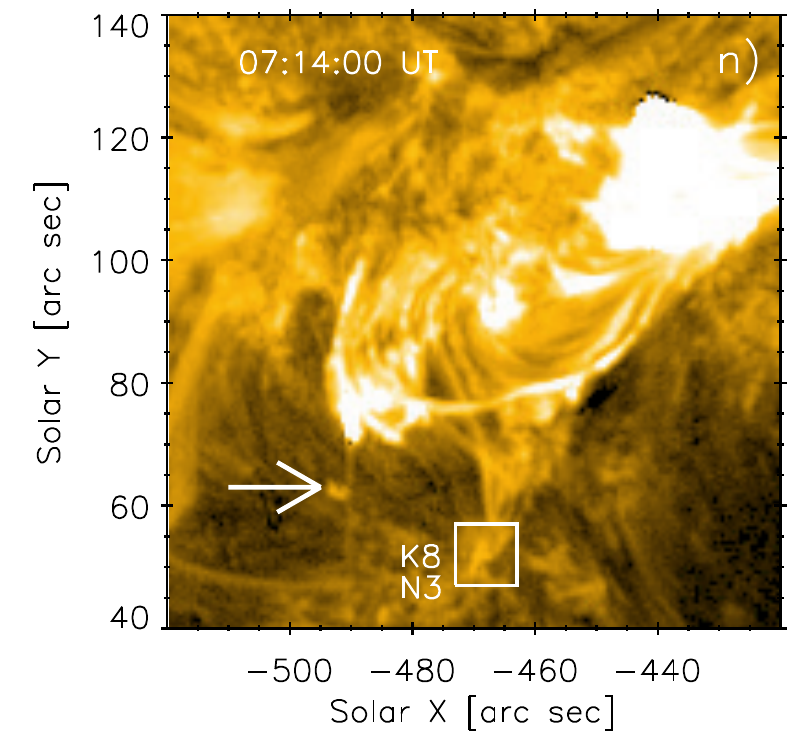} 
\includegraphics[width=4.4cm, clip, viewport=78 0 378 348]{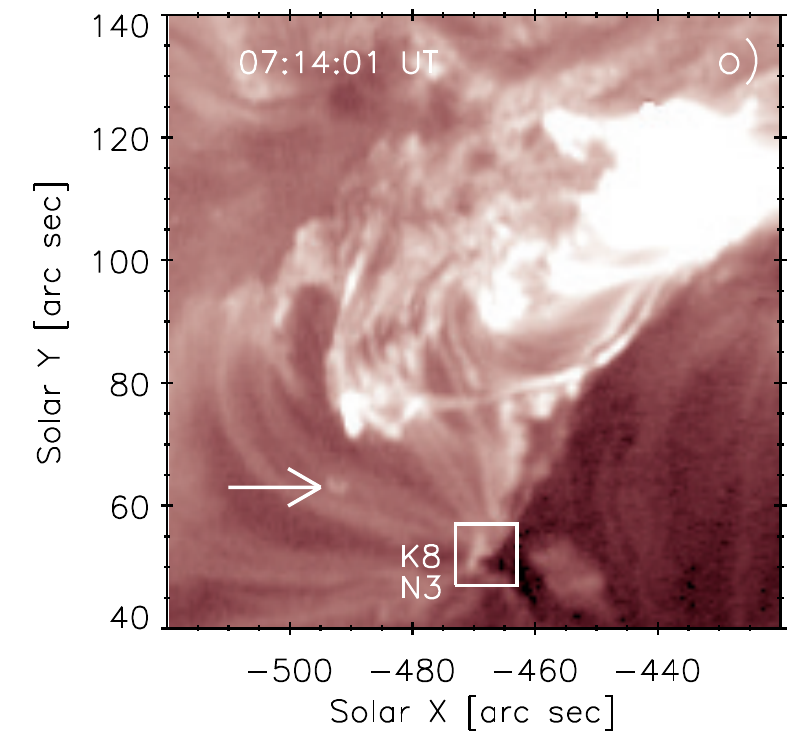} 
\caption{Evolution of the swirl in three SDO/AIA filters, (131, 171, and 211 \AA). The white rectangle shows the position of K8 in the spot N3. The black circular arrow in (e) shows the direction of plasma rotation in the swirl; the  white arrows in (d)--(f) show a jet of plasma towards N3; the arrows in (h)--(i) and (k)--(l) point to the ray-like structure moving away from the swirling plasma. The last bubble of plasma is shown in (n)--(o). The temporal evolution of the swirl (06:58--07:15 UT) in SDO/AIA filters 131 \AA, 171 \AA,~and 211 \AA~is available as an online movie (swirl\_131\_171\_211.mpg).}
\label{fig_swirl}
\end{figure*}

Finally, Figs.~\ref{fig_flare_1}h and~\ref{fig_flare_1}i show an additional feature of interest present during this flare: within the white circle there is a ray-like system of bright EUV features. The white circle is centred at [$-490$\arcsec, $70$\arcsec], near the top of the elbow of the S, and later the arcade of flare loops is seen in the flare core. These ray-like features are bubbles of plasma moving away from the top part of the loop. Figure~\ref{fig_swirl} (and the movie associated with it) shows this phenomenon in detail. After 06:59 UT we observed the formation of a plasma swirl north of N3 and at the elbow of S (Fig.~\ref{fig_swirl}a--d).  The swirling motion of various plasma blobs was visible in AIA 171 \AA, 211 \AA, but also 131 \AA~starting at about 07:00 UT. The visibility of the swirl in these three filters indicates that it likely originated at temperatures below 1 MK \citep{ODwyer2010,DelZanna2013}. The sense of the plasma rotation in a swirl is indicated by a black arrow in Fig.~\ref{fig_swirl}e. While the plasma was seen to rotate, the centre of a swirl was slightly moving to the south-east and we could observe several short bright jets towards  N3 (e.g. Fig.~\ref{fig_swirl}e). 
About 07:08 UT several ray-like structures and blobs of plasma were seen to separate from the area of fading swirl (Fig.~\ref{fig_swirl}h, k, n; i, l, o, arrows). The swirl completely fades away at about 07:14 UT.

%

\section{Analysis of UV/EUV light curves from individual flare kernels}
\label{sec_uv_lc}

Since the radio observations of interest, including the SPDBs and type III bursts, do not contain positional information, the only possibility to connect them to various flare phenomena observed in UV and EUV is by temporal correlation. This is the focus of the remainder of this section.

The first flare kernels K1 and K2 appeared two minutes before the onset of the GOES C8.7 SXR flare. The radio burst activity within 600--5000 MHz was observed few minutes later, during its impulsive phase and about its maximum (Fig.~\ref{fig_X-ray}b), but not before 06:57 UT (i.e. not before the start of the impulsive phase).

The time interval 06:57:48--06:58:46 UT, during which groups III-A and III-B were observed in 2000--5000 MHz, as well as group of SPDBs-1 in 800--2000 MHz, coincides in time with the appearance of two sets of flare kernels: those to the south of N3 (K3, K4, K6), and those located along the ribbons (K10--K13); Fig.~\ref{fig_mag}c). At the location of K10--K12 we observed intensive HXR sources (Figs.~\ref{fig_reco}c,~\ref{fig_flare_1}e) detected during time intervals 06:57:40--06:58:33 UT and 06:58:34--06:59:28 UT. The brightest EUV kernel on the south of N3 was K3 where we did not observe any HXR source. If any weak HXR source was excited there, RHESSI likely could not resolve it due to its low dynamic range. Thus, the intensive RHESSI HXR sources were located only along parallel ribbons in the flare core. 

To find  any temporal association of detected radio burst activity with the processes producing flare kernels, we followed the  appearance of flare kernels in the UV SDO/AIA 1600 \AA~filter. The emission observed in this filter during the flares \citep{Simoes2019} comes mainly from flare ribbons with little or no contribution from flare loops. Therefore, we followed the appearance of flare kernels from the beginning of the flare to approximately its maximum because we looked for their association with radio bursts observed from 06:57 UT to 07:03 UT (Fig.~\ref{fig_X-ray}b).

Then we constructed LCs from the kernels in two SDO/AIA filters: 1600 \AA~and 304 \AA. These two filters observe emission from the transition region or upper chromosphere, but also with other contributions \citep[see][for details]{ODwyer2010,Simoes2019}.
The LCs were constructed from regions of approximately $11$\arcsec~$\times$ $11$\arcsec\ in size, centred on individual kernels (Fig.~\ref{fig_mag}c). This size was chosen to encompass the extent of observed flare kernels in 1600 \AA~passband (see movie associated with Fig.~\ref{fig_flare_1}). The temporal evolution of the normalized UV/EUV LCs was compared with that of the normalized radio flux at a frequency of 1190 MHz where narrow-band type III and all three groups of interesting SPDBs were registered. 

\begin{table}[]
\caption{Times of individual peaks at 1190 MHz rounded to seconds (see Figs.~\ref{fig_lc} and~\ref{fig_lc_12055}).}
    \centering
    \begin{tabular}{c c c}
    \hline\hline
    Peak label  &  & Time\\
    & & $[$UT$]$\\
    \hline
      1   &  & 06:58:03\\
      III &  & 06:59:36\\
      2   &  & 07:01:22\\
      3   &  & 07:02:15\\
      \hline
    \end{tabular}
    \label{table:2}
\end{table}

\begin{figure*}
\centering
\includegraphics[width=8.6cm, clip, viewport=25 48 490 328]{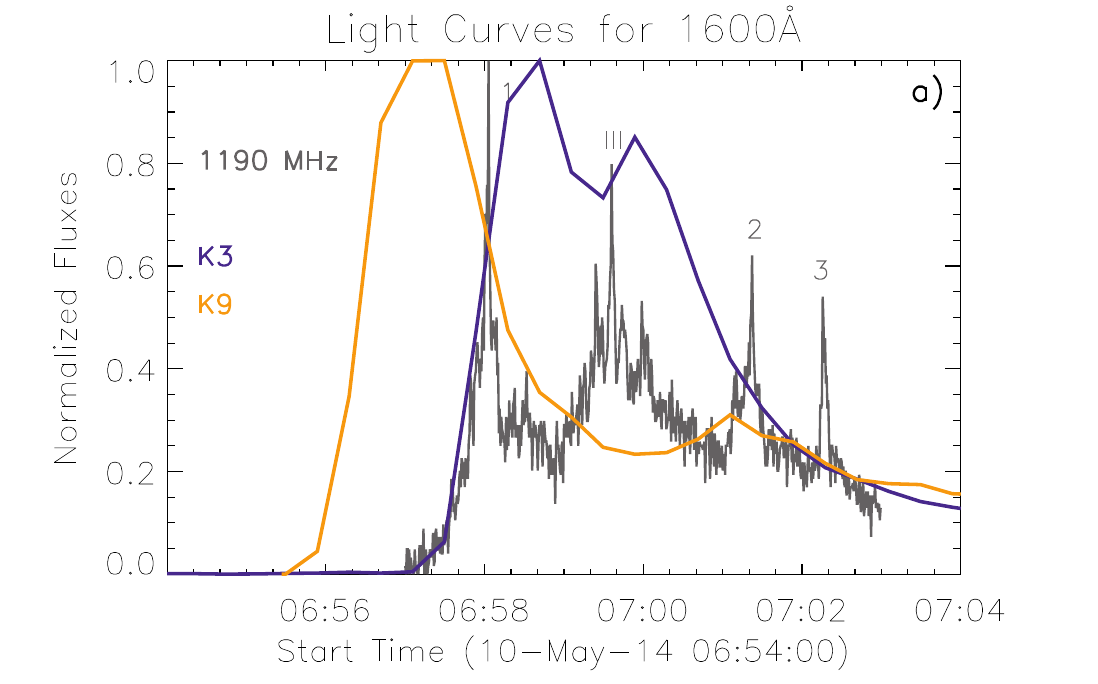}
\includegraphics[width=7.6cm, clip, viewport=80 48 490 328]{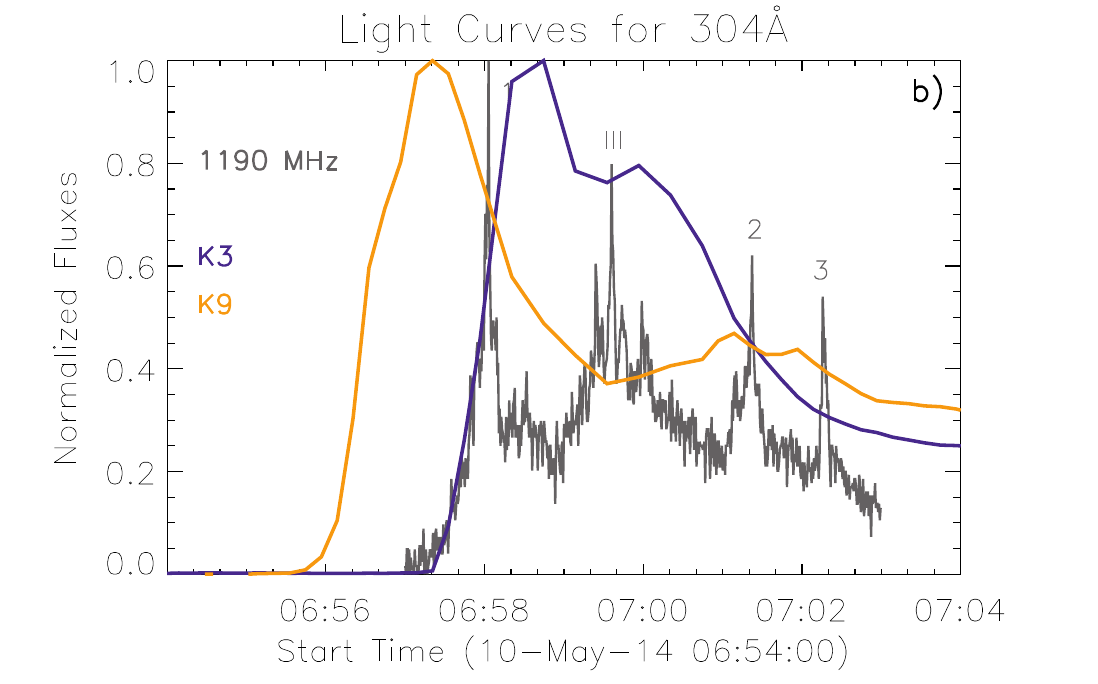}

\includegraphics[width=8.6cm, clip, viewport=25 48 490 300]{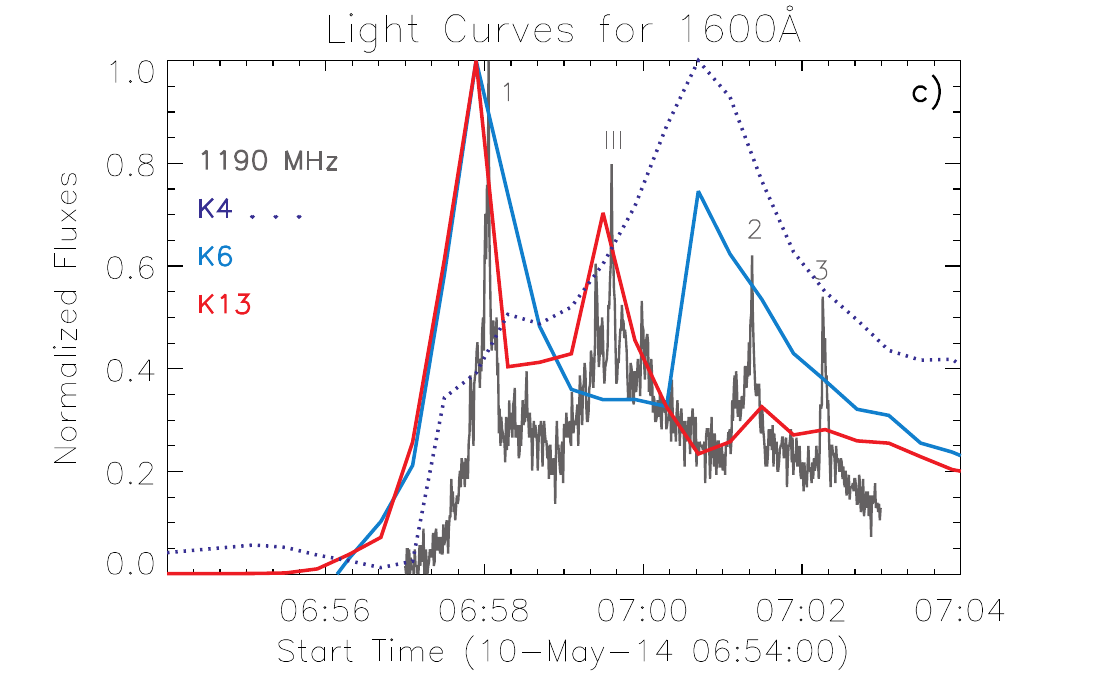}
\includegraphics[width=7.6cm, clip, viewport=80 48 490 300]{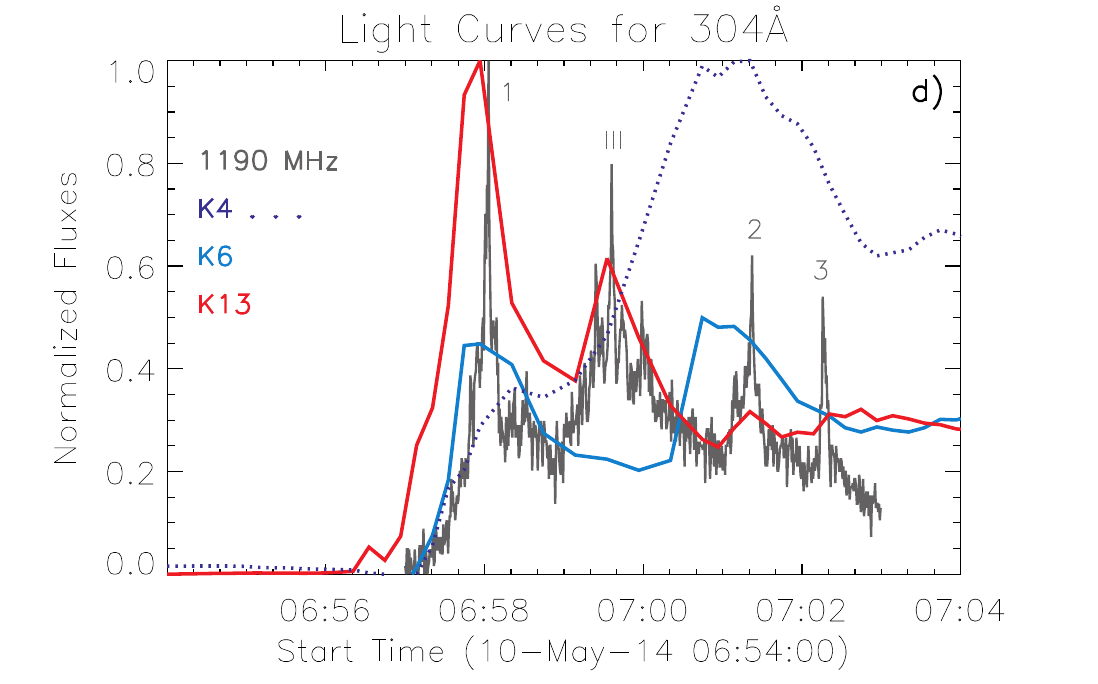}

\includegraphics[width=8.6cm, clip, viewport=25 48 490 300]{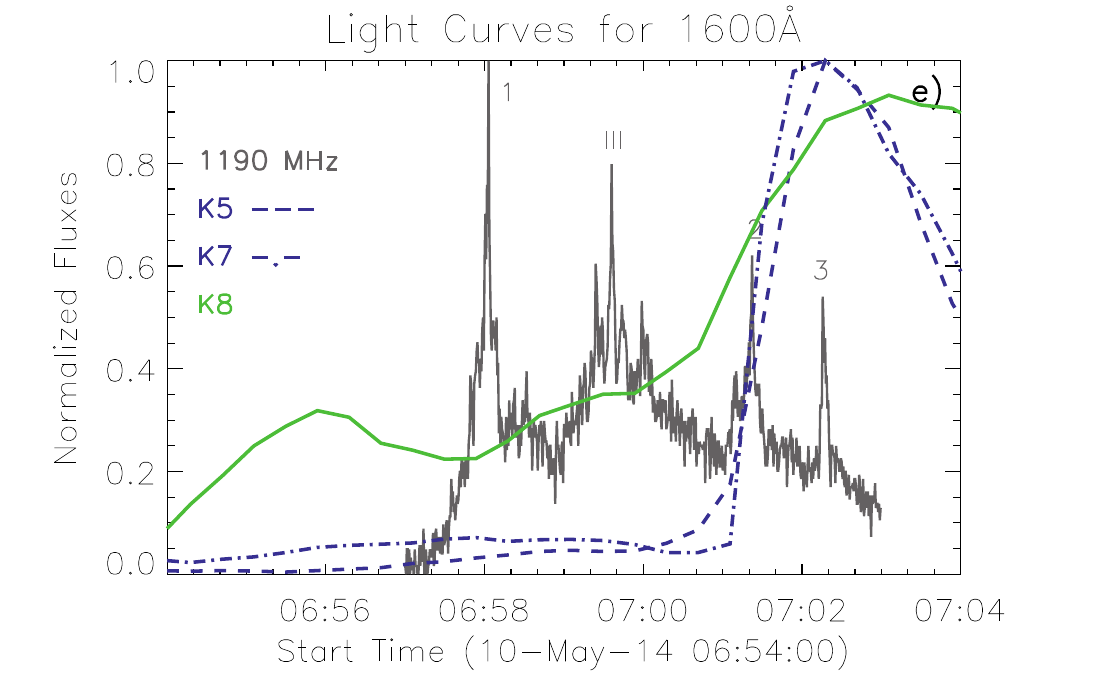}
\includegraphics[width=7.6cm, clip, viewport=80 48 490 300]{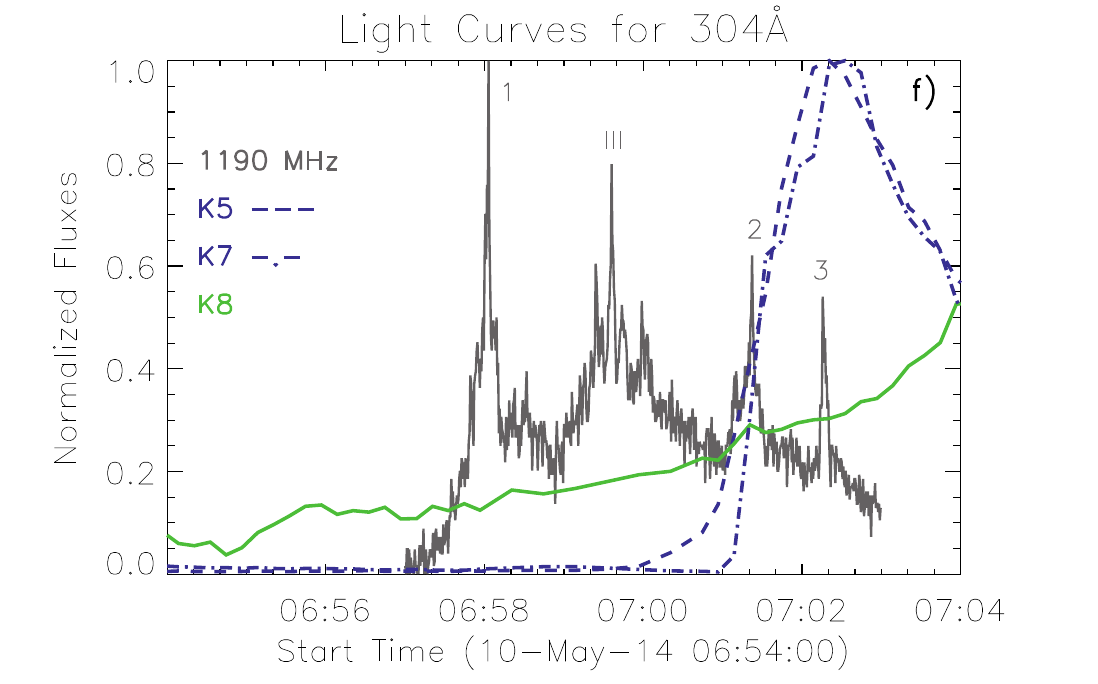}

\includegraphics[width=8.6cm, clip, viewport=25  5 490 300]{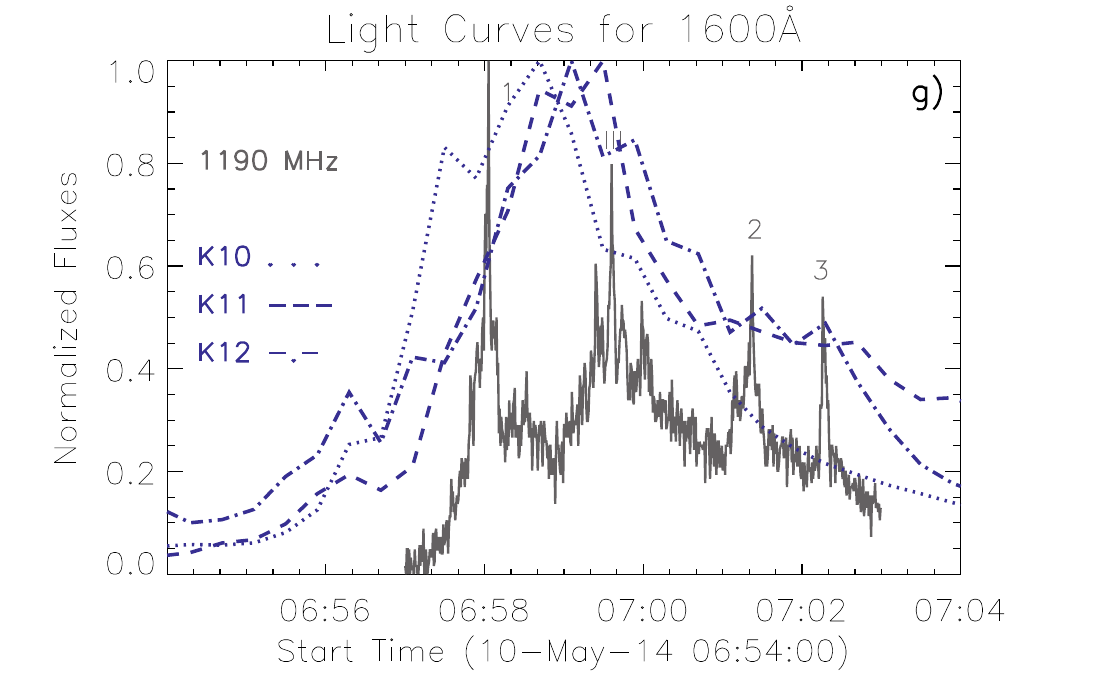}
\includegraphics[width=7.6cm, clip, viewport=80  5 490 300]{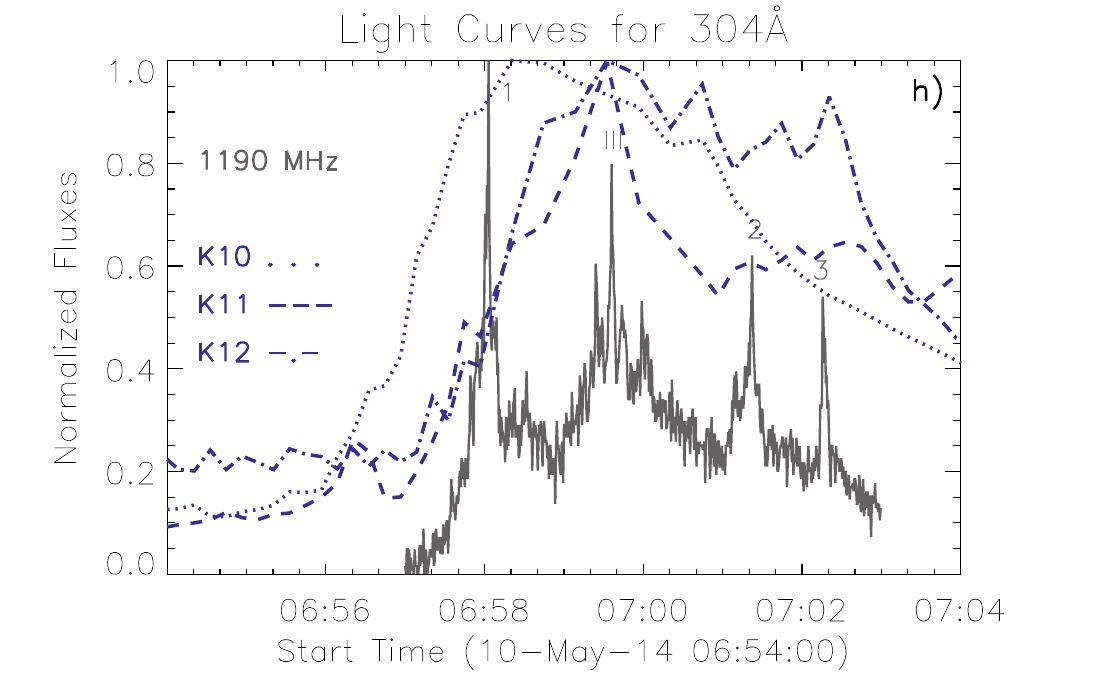}
\caption{Comparison of temporal evolution of UV/EUV light curves from flare kernels K3--K13 (see Fig.~\ref{fig_mag}c) with the normalized radio flux at 1190 MHz (grey). The radio peaks corresponding to groups of SPDBs are labelled  1, 2, and 3, and III denotes narrow-band type III bursts. (a) For UV filter 1600 \AA~and (b) for EUV filter 304 \AA. The individual kernels are indicated by different colours or linestyles.}
    \label{fig_lc}
\end{figure*}

The result of the comparison is shown in Fig.~\ref{fig_lc}. This figure has two columns, one for each filter (1600 \AA~and 304 \AA), and the panels are grouped according to the relation of kernels to the flare evolution. 
The cut through the radio spectrum is plotted in grey, showing four dominant peaks. They are marked 1, 2, 3, and III, and their times are summarized in Table~\ref{table:2}. The numbers 1--3 mark the groups of SPDBs and the peak labelled III marks the group of narrow-band type III bursts superimposed on the low-frequency part of the enhanced continuum emission at 800--2000 MHz (Fig.~\ref{fig_rspec} and Sect. \ref{sec_radio}). 
The relatively low cadence of SDO/AIA images, 24 s for the 1600 \AA~filter images and 12 s for the 304 \AA~filter, does not make it easy to directly compare the UV/EUV LCs with radio emission, which has much better time resolution (0.01 s). Nevertheless, distinct peaks in the  UV/EUV LCs can be still identified, allowing us to compare their temporal association with radio emission at 1190 MHz. 

%

\subsection{Kernels K3 and K9}
\label{sec_K3_K9}

We start with a comparison of the UV/EUV LCs from kernels K3 and K9 (Fig.~\ref{fig_lc}a--b). The southernmost kernel K3 as well as   kernel K9 located at the edge of the half dome brighten approximately simultaneously as the flare loops of S reach the half dome (Sect. \ref{sec_flare_core}, Fig.~\ref{fig_mag}c and Fig.~\ref{fig_reco}c).
The violet LCs in 1600 \AA~and 304 \AA~belonging to K3 both show two peaks during the time interval of the observed radio bursts. The first and the highest peak appeared at about 06:58:40 UT and the second   occurred slightly before 07:00:00 UT. The ascending parts of these UV/EUV LCs follow the rise of   radio peak 1 (group of SPDBs-1), but the maximum is delayed by about 40 s relative to this peak. The second peak in the UV/EUV LC from K3 is delayed by about 20 s relative to the narrow-band type III bursts seen over the enhanced continuum emission at low frequencies (Fig.~\ref{fig_rspec}a). For the UV/EUV LCs from K3, no other peaks were observed that could be associated with SPDBs-2 and SPDBs-3.

The orange LCs belong  to   kernel K9 (Fig.~\ref{fig_reco}c), which was located at the edge of the half dome ribbon (Fig.~\ref{fig_mag}d). The maxima of this UV/EUV LCs occurred at 06:57:00--06:57:40 UT (with a more pronounced maximum in 304 \AA); that is, about the time when loops of PRH moved through this region and NRH reached the sunspot N2. Orange LCs from K9 also have  small local peaks slightly before SPDBs-2 (1600 \AA, 304 \AA) and SPDBs-3 (304 \AA). 

\subsection{Kernels K4, K6, and K13}
\label{sec_K4_K6_K13}

We next investigated the kernels K4, K6, and K13. These kernels appeared together with K3 but were fainter (see Sect.~\ref{sec_K3}). K6 was located at the southern end of the half dome, while K4 was located between it and K3 (see Figs. \ref{fig_reco}c and \ref{fig_hl}a), while kernel K13 was located close to the N2 sunspot (Fig.~\ref{fig_mag}d) at the end of  NRH (i.e. conjugate location to kernel K9; see Sect. \ref{sec_K3_K9}). Their respective UV/EUV LCs are shown in Fig.~\ref{fig_lc}c and d.

The UV/EUV LC from K4 (dotted line) shows two local peaks. In both passbands, the first peak is smaller and occurs slightly later than radio peak 1. The second peaks are wide and occur at 07:00:40--07:01:20 UT, between radio peaks III and 2. Overall, there is little correlation of the K4 LCs with the radio flux at 1190 MHz.

The LCs from kernel K6 (light blue) also show two peaks. The first ones have a steep rise and occur slightly before   radio peak 1. The second ones occur at about 07:00:40 UT before radio peak 2. It is seen that the second peaks are not correlated with the radio emission.

The red curves in Fig.~\ref{fig_lc}c--d belong to kernel K13. These LCs show three local peaks: the main one which coincides within few seconds with the radio peak 1, the second one which occurred about the time of narrow-band type III bursts, and the last one that occurred about the time of  radio peak 2. At the time of radio peak 3 these LCs show enhanced emission only, but there is no discernible peak. These LCs from kernel K13 follow best the peaks seen in radio emission at 1190 MHz, showing distinct similarity to the radio flux during peaks 1, III, and 2.

\subsection{Other kernels: K5, K7, K8, and K10--K12}
\label{sec_other_kernels}

Kernels K5 and K7 are located outside of the half dome within the system of extended ribbons (Fig.~\ref{fig_flare_1}i), while kernel K8 is located approximately at the centre  of the sunspot N3 (Fig. \ref{fig_mag}). Their respective LCs are shown in Fig.~\ref{fig_lc}e--f. The UV/EUV LCs from K5 and K7 start to rise steeply at about 07:01:00 UT reaching their maxima between 07:02:00--07:02:20 UT, covering radio peak 3, and then they steeply decrease. This indicates that the SPDB-3 could be related to the reconnection at the half dome, and establishing new connections to the far kernels K5 and K7.
The UV/EUV LCs from K8 (green) increase in both wavelengths quite soon, at $\approx$ 06:55 UT, and their evolution is not obviously correlated to the radio emission. 

Finally, Fig.~\ref{fig_lc}g and h shows the UV/EUV LCs from three kernels, K10 (dotted), K11 (dashed), and K12 (dash-dotted), located along parallel ribbons below S (Fig.~\ref{fig_mag}c). The UV LCs from these regions started to rise earlier (at about 06:56 UT) than any radio bursts occurred. The maxima of these LCs occurred between 06:58--07:00 UT in agreement with an enhancement of HXR flux in 12--50 keV (Fig.~\ref{fig_rspec}d). 
There is no obvious association of the K10--K12 LCs with the radio flux at 1190 MHz, perhaps aside from the coincidental occurrence of small peaks in 304 \AA~LC of K12 with radio peaks III and 3. However, these peaks are relatively small, and there are other similar peaks that are not associated with radio emission. Therefore, the locations within the straight ribbons are likely not related to SPDBs. These kernels K10--K12 could still be associated with enhanced continuum and groups III-A and III-B (see Sect. \ref{sec_interp_other_bursts}).

%
\begin{figure*}
\centering
\includegraphics[width=9cm, clip, viewport= 2 0 378 348]{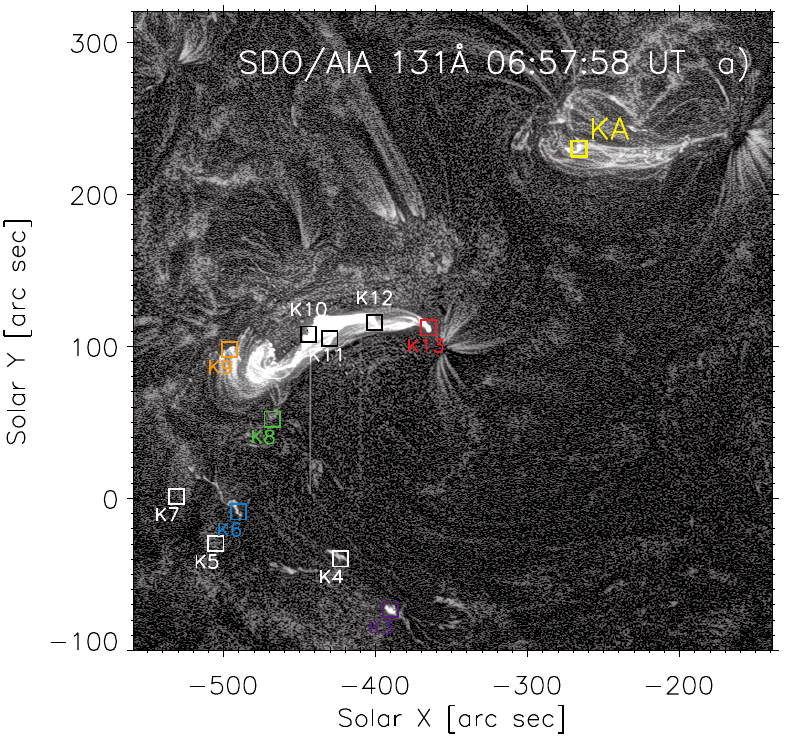} 
\includegraphics[width=7.6cm, clip, viewport=60 0 378 348]{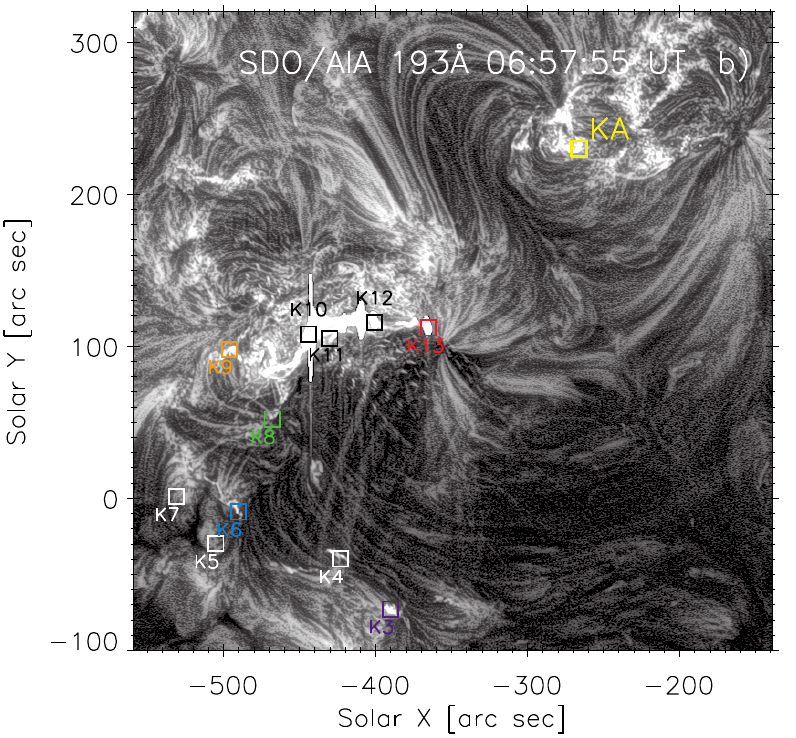} 
 \caption{View of flaring active region NOAA 12\,056 and neighbouring NOAA 12\,055. (a) 131 \AA~filter showing bright loops in NOAA 12\,055 next to  kernel KA (yellow square). (b) 193 \AA~filter showing interconnecting loops between the two active regions between flare kernel K13 (red square) and kernel KA (yellow square). Kernels K3--K13 from C8.7 are depicted in both panels as square regions. The yellow square marked KA is the region from which light curves are shown in Fig.~\ref{fig_lc_12055}. The temporal evolution of the C8.7 flare (06:45--07:45 UT) in SDO/AIA filters 131\AA~and 193\AA, and its connection to the neighbouring active region is available as an online movie (mgn\_131\_193.mpg).}
   \label{fig_12055}
 \end{figure*}
 
%
\begin{figure*}
\centering
\includegraphics[width=8.5cm,clip, viewport=25 5 490 328]{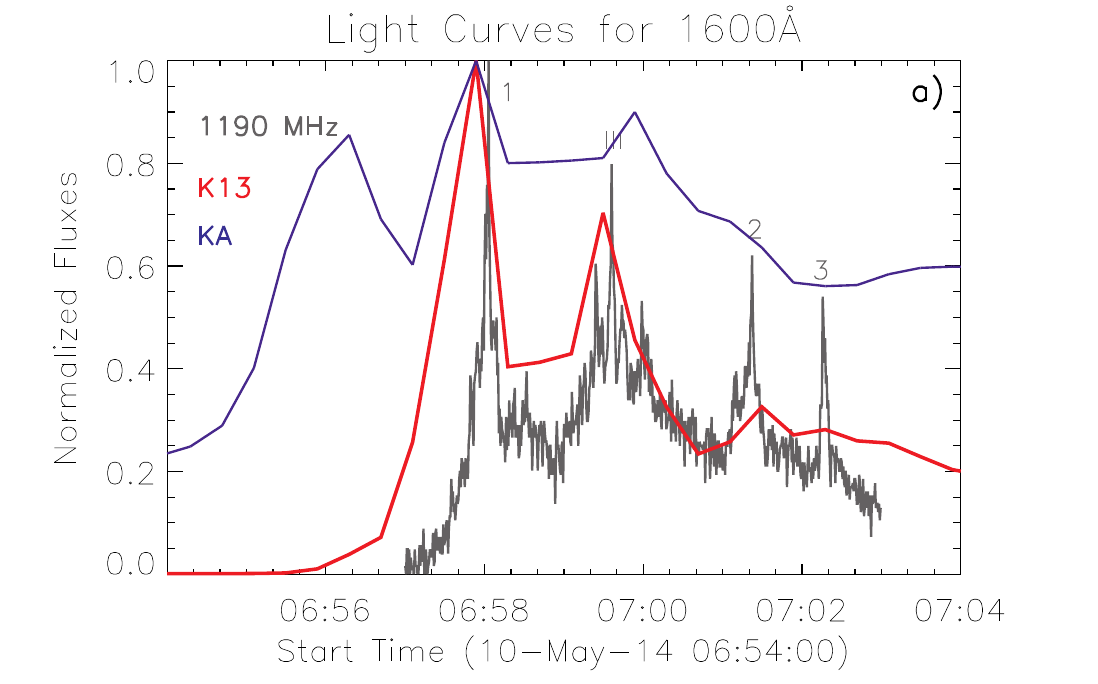}
\includegraphics[width=8.5cm,clip, viewport=25 5 490 328]{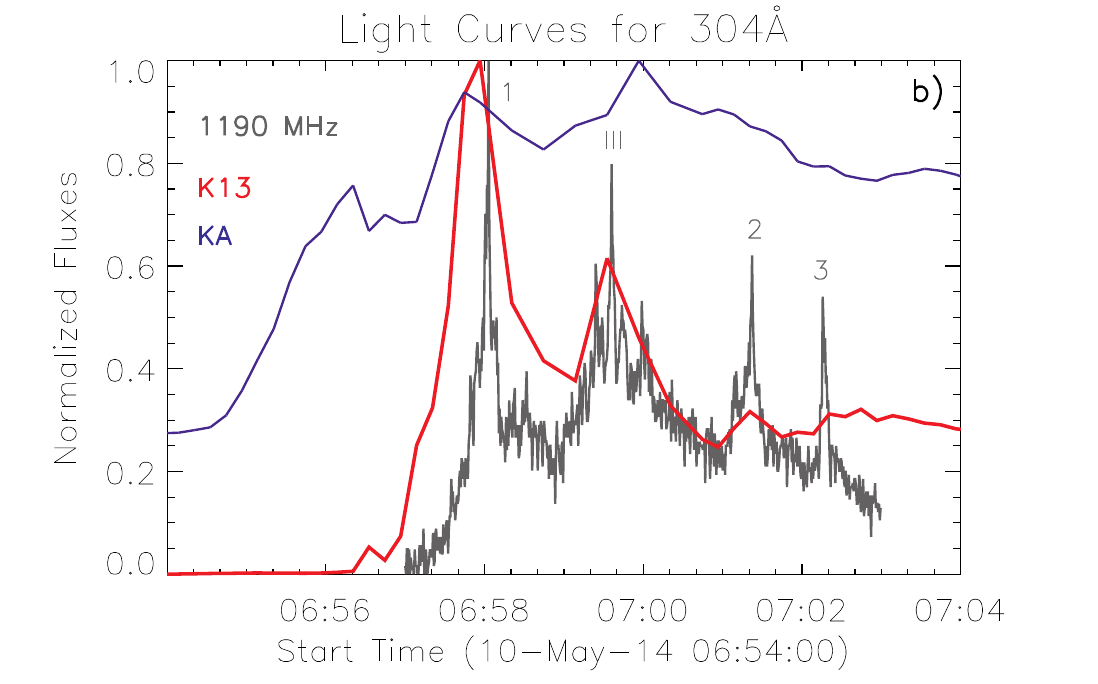}
\caption{Comparison of light curves from kernel K13 located in active region NOAA 12\,056 and kernel KA located in NOAA 12\,055. (a) Light curves from the  1600 \AA~filter and (b) for the  304 \AA~filter. K13 is in red and KA in dark blue. The normalized radio flux at 1190 MHz is in grey.}
    \label{fig_lc_12055}
\end{figure*}

\subsection{Kernel KA in neighbouring active region NOAA 12\,055}
\label{sec_KA}

As mentioned in Sect.~\ref{sec_flare}), during the time of the C8.7 flare in NOAA 12\,056 some hot loops also appeared  in the neighbouring active region NOAA 12\,055 (Fig.~\ref{fig_12055} and the associated animation). Therefore, we investigated the UV kernels in 1600 \AA~appearing there during the time of the observed radio bursts, and also constructed the respective UV/EUV LCs. 

Only one kernel, KA, showed a possible relation to the observed radio bursts (Fig.~\ref{fig_12055}). 
The UV/EUV LCs from KA are plotted in blue in    Fig.~\ref{fig_lc_12055}, together with the radio emission at 1190 MHz. For comparison, we also added the UV/EUV LCs from K13 in NOAA 12\,056. In both wavelengths, KA shows peaks about the radio peaks 1 and III similarly to K13 (red), although the KA LCs are not as similar to the radio flux at 1190 MHz as the K13 LCs. Even so, at least partial similarity in evolution of LCs from KA can indicate partial association with observed radio bursts. 
Although no direct connectivity between K13 and KA is visible in AIA observations, some low-lying magnetic connections (loops) between NOAA 12\,055 and 12\,056 are present in AIA 193 \AA~(Fig.~\ref{fig_12055}b). Therefore, a possible large-scale connection between K13 and KA is still possible.

\section{Interpretation and discussion}
\label{sec_interp}

\subsection{Large-scale magnetic configuration}
\label{sec_interp_global}
Based on the previous, we propose that interaction of the S with the half dome structure resembles a scenario of 3D fan-spine magnetic reconnection introduced by \cite{Pontin2013}. 
However, the presence of a 3D coronal null point necessitates the presence of a full dome constituting a fan surface, as well as an isolated spine field line \citep{Priest2016,Li2021}. During a flare, a circular ribbon is expected to be observed together with a brightening of a remote ribbon located at the outer spine footpoint \citep[e.g.][]{Masson2009, Reid2012, Sun2013, Prasad2018}. Instead, here we have the half dome configuration and not a full dome. This means we are likely dealing with a quasi-separatrix layer (QSL) \citep{Demoulin1995, Demoulin2006} instead of a full fan separatrix surface. Such quasi-separatrix layers are favourable locations where the strong electric currents can accumulate allowing   3D magnetic reconnection to occur \citep{Demoulin1995, Aulanier2005, Aulanier2006, Pontin2013, Priest2023}. Real 3D flaring magnetic configurations in the corona can possess several such QSLs which cross themselves in quasi-separators \citep{Priest2023} that have a structure of HFT \citep[][]{Titov2002}. Three-dimensional magnetic reconnection then occurs preferably near null point, separator, or HFT \citep[see][and references therein]{Priest2023}.

The half dome we study represents the QSL that shows dramatic change in magnetic connectivity from the side of the N3 sunspot towards the semi-circle of positive network polarity (Fig.~\ref{fig_mag}a, b). Due to the strongly localized concentration of negative magnetic field in sunspot N3, the configuration looks to have a null point with an analogue to the spine field line. As noted above, the observed magnetic configuration is not fully identical to the simple model of \citet{Pontin2013} as there is no null point, but rather a HFT. Nevertheless, the model of \citet{Pontin2013} can be used for simplified description of the observed flare and helps us to explain the association of individual kernels with detected radio bursts.

Figure~\ref{fig_mag}d shows locations of flare kernels that can be compared with footpoints shown in Figs. 5 and 6 of the model by \citet{Pontin2013}. It shows kernels K3 and K4 to the south of N3, then K6 and  K9 located at the faint semi-circular H$\alpha$ ribbon, and K8 and K13 located at negative magnetic fields of N3 and close to N2. We propose that K8 and K13 are located at the inner and outer footpoints, respectively, of the spine-analogue field line. Thus, we expect the HFT to be situated above the N3 sunspot, close to the observed plasma swirl (Fig.~\ref{fig_swirl}e). The loops of the rising S we consider as field lines of ambient magnetic field that underwent HFT magnetic reconnection associated with the half dome structure. We think that this process of 3D magnetic reconnection generated the observed radio emission in 600--5000 MHz. A sketch explaining this in three steps is shown in  Fig.~\ref{fig_schema}, and is described in   Sect. \ref{sec_interp_evol}. 

\subsection{Evolution of magnetic connectivity during the flare}
\label{sec_interp_evol}

\begin{figure}
\centering
\includegraphics[width=8.0cm,clip, viewport= 2 40 340 350]{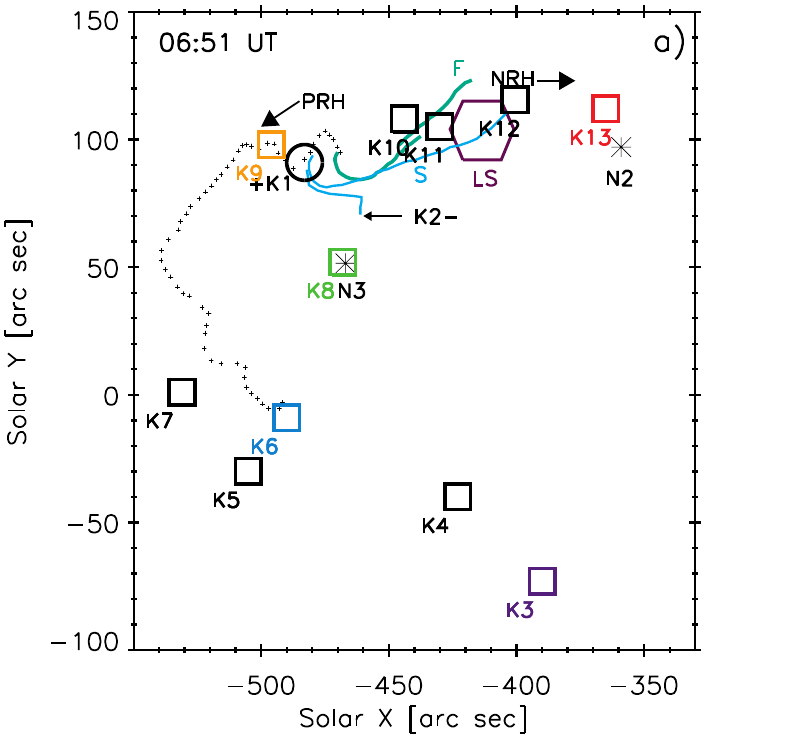}
\includegraphics[width=8.0cm,clip, viewport= 2 40 340 350]{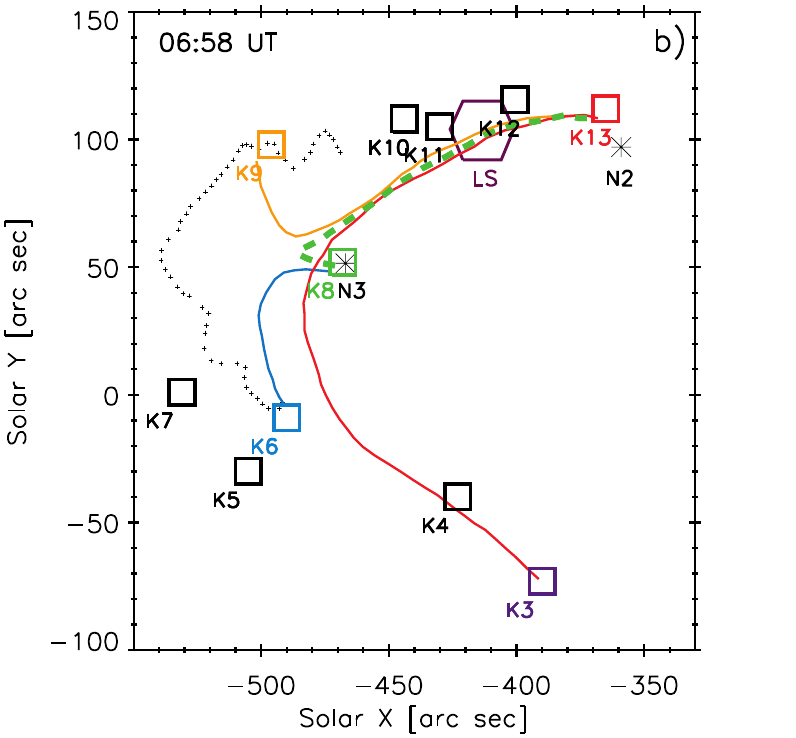}
\includegraphics[width=8.0cm,clip, viewport= 2 0 340 350]{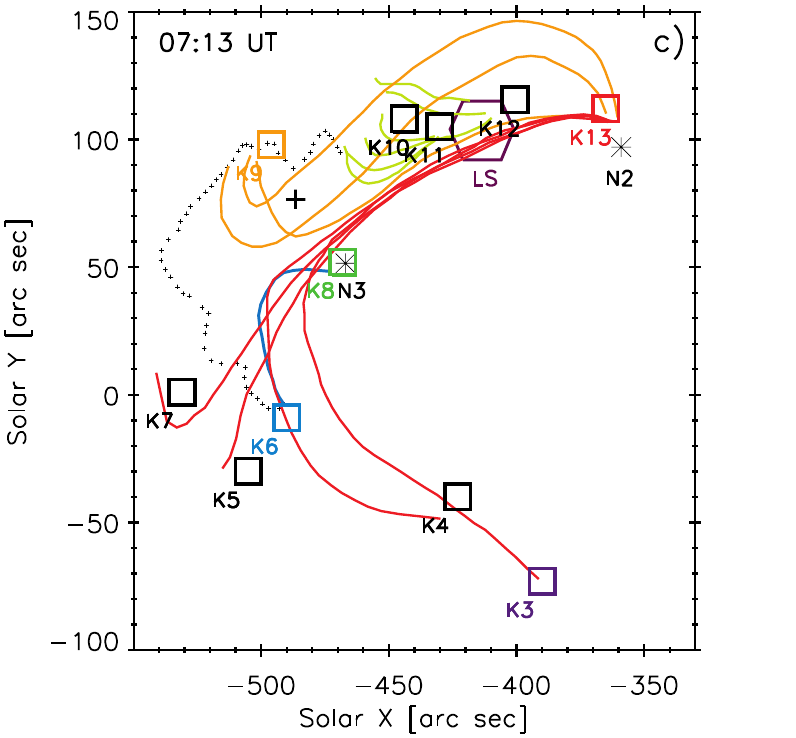}
\caption{Schema of the flare and evolution of individual magnetic connections during it. The asterisks indicate the positions of sunspots N2 and N3. Tiny black crosses denote the half dome. Individual kernels are marked by squares, while coloured lines stand for different magnetic connections. See text of Sect. \ref{sec_interp_evol} for details on individual structures.}
 \label{fig_schema}
\end{figure}

Based on the observed evolution, we produced a schematic scenario for the evolution of the flare including the observed magnetic connectivity and associated changes. This scenario is summarized in Fig.~\ref{fig_schema}.

Figure~\ref{fig_schema}a shows the situation at the beginning of the flare (06:51 UT). The S that formed over  filament F (teal), is shown as a light blue loop connecting K1 and K12. At this time S is located within the core of active region NOAA 12\,055, outside and to the west of the half dome structure (Fig.~\ref{fig_flare_1}a). 
The interaction between S and the half dome started by  the tiny loops K1-K2 (shown as a short light blue loop connecting K1 and K2) slipping towards N3. Furthermore, S interacted with overlying loops L1 and L2 (Fig.~\ref{fig_flare_1}b), and then it slipped through the position of K1 (Fig.~\ref{fig_reco}b, c). 
Simultaneously the hooked ribbons of S were shifting to new positions (black arrows in Fig. \ref{fig_schema}a, labelled PRH and NRH). This  shift of the erupting flux rope footpoints was described by \citet{Aulanier2019} and later observed by multiple authors \citep{Zemanova2019, Lorincik2019, Dudik2019, ChenH2019, Peng2022, Gou2023, Guo2024}. The start of interaction between S and the half dome is co-temporal with rise in the GOES SXR flux at about 06:50 UT (Fig.~\ref{fig_X-ray}a), but no radio emission at 600-5000 MHz has been observed yet.

After 06:57 UT, that is, during the impulsive phase, the slipping and growing S reached the position of K9 (Fig.~\ref{fig_reco}c), represented by orange loop connecting K9 and K13 in Fig.~\ref{fig_schema}b. At this time  NRH of the growing flux rope S has already reached K13. This means that some of the hot loops of S were approaching the spine line analogue (green dashed line). As the orange-type field lines entered into the reconnection area near the HFT (located above N3), the continuous change in connectivity led to the appearance of new large-scale magnetic connections and flare kernels K3, K4, K6. The new flare loops connecting these kernels are also shown in Fig.~\ref{fig_schema}b. The red loop now connects the  negative K13 with the positive kernel K3 in the south. This process was associated with radio emission observed in 600--2000 MHz.
In Sect. \ref{sec_uv_lc} we show that K13 brightened almost simultaneously with maximum of radio flux representing the group of SPDBs-1 (Fig.~\ref{fig_lc}c and d). Moreover, the rise in radio peak 1 was also nicely copied by rise in LCs from K3 (Fig.~\ref{fig_lc}a and b). The delayed maximum of LCs in K3 can be explained by local plasma heating, followed by its evaporation (Fig.~\ref{fig_hl}a). We propose that the group of SPDBs-1 was associated with this step in 3D magnetic reconnection.

At about 07:00 UT the plasma swirl was also observed in the elbow of S, and north of K8 in N3, that is, within the northern part of the half dome. Both the location and time of the occurrence suggest that the swirl is also related to the ongoing reconnection of S at the HFT. The swirl appeared shortly after SPDBs-1, by less than about two minutes. Although SPDBs-2 and SPDBs-3 occurred shortly after the onset of the swirl, we cannot relate the swirl to the SPDBs-1 directly.

Figure~\ref{fig_schema}c shows the situation about the maximum of the flare, when both SPDBs-2 and SPDBs-3 were observed. We propose that these two groups were associated with subsequent episodes of HFT reconnections occurring after 07:00 UT. At this phase of the flare, kernels K5 and  K7 appeared, and other large-scale flare loops were observed. These are shown in red in Fig.~\ref{fig_schema}c. Again, we support our explanation by analysis of LCs from K5 and K7 which started to rise suddenly about the time of SPDBs-2 (Figs.~\ref{fig_lc}e--f). After 07:03 UT, no further radio bursts were observed at 600--5000 MHz. 

We recall that the groups of SPDBs-1, SPDBs-2, and SPDBs-3 did not follow each other directly. Between the first and the second group the narrow-band type III bursts (Table~\ref{table:1}) occurred together with continuum emission (Fig.~\ref{fig_rspec}). We suggest that these type III bursts are also connected to HFT reconnection because the LCs from K13 (at outer footpoint of {the spine line analogue}) also show peaks that correlate with the radio flux during the group of type III bursts.

\subsection{Two-ribbon flare with no eruption}
\label{sec_interp_no_eruption}

The flare as observed in EUV and H$\alpha$ shows classic characteristics of a two-ribbon eruptive flare. It starts at a location of  filament F, is characterized by a growing S \citep[or `hot channel'; see e.g.][]{Zhang2012,Cheng2013,Cheng2015,Dudik2014,Dudik2016,Hernandedz-Perez2019,Liu2022,Zhang2023}, and is associated with a pair of hook-shaped ribbons \citep[see e.g.][]{Aulanier2012_I,Janvier2013_III,Janvier2014,Zhao2016,Lorincik2019}. However,  there was neither a discernible eruption of the sigmoid (hot channel) nor a CME, according to the LASCO CDAW catalogue.\footnote[5]{\url{https://cdaw.gsfc.nasa.gov/CME_list/UNIVERSAL_ver1/2014_05/univ2014_05.html}}

A possible explanation can be that the S indeed grows unstable, but its eruption is prevented by the reconnections at the HFT (Sect. \ref{sec_interp_evol}). This is supported by the fact that the associated ribbon hook, PRH, which encircles the footpoints of the evolving magnetic flux rope \citep[see e.g.][]{Aulanier2019}, extends until it reaches the half dome near the location of K9 at about 06:57 UT (Sect. \ref{sec_flare_core}).
Essentially, the eruption is `shredded' by these reconnections at the HFT and the associated half dome. 
A shredding scenario   of an eruptive flux rope at the HFT was already described by \citet{Chintzoglou2017}. 
An interesting feature of our flare is that such a process is related to the generation of SPDBs.

\subsection{Radio bursts III-A and III-B}
\label{sec_interp_other_bursts}

In addition to SPDBs 1--3 and the group type III bursts studied at 1190 MHz, other radio bursts were observed at frequencies above 2 GHz~(Table \ref{table:1}). We argue here that these were not produced in the HFT reconnection process.

From 06:56 UT, a pair of J-shaped ribbons was observed in the flare core (Fig.~\ref{fig_mag})c.  Continuum emission in 2000-5000 MHz range was observed from 06:57:40 UT and the groups of radio bursts III-A and III-B were detected few seconds later (Table~\ref{table:1}). The group SPDBs-1 (700--1300 MHz) and the group III-A (4100-5000 MHz) started simultaneously, while the group III-B (4100-4800 MHz) followed after 32 s. 
Radio bursts at these relatively high frequencies were likely associated with kernels K10--K12. Their LCs started to rise faster at about 06:56 UT, with the maxima dispersed between 06:58--07:00 UT (Figs.~\ref{fig_lc}g--h). The HXR emission also started to rise before 06:57 UT (Fig.~\ref{fig_rspec}d) and all strong HXR sources (in 25--50 keV) were observed along the parallel ribbons only (Fig.~\ref{fig_flare_1}e). Therefore, the groups III-A and III-B of narrow-band type-III bursts seen over continuum were associated with growth of the S and formation of the arcade of flare loops below S (yellow loops in Fig.~\ref{fig_schema}c). Considering the plasma mechanism for generating these groups of bursts, the corresponding plasma densities are $2$--$3\times10^{11}$\,cm$^{-3}$ (fundamental) or $0.5$--$0.75\times10^{11}$\,cm$^{-3}$ (harmonic). Such plasma densities are in agreement with typical electron densities of flare loops and their footpoints found from EUV spectroscopy \citep{Milligan2011, Graham2015}.

\subsection{Duration of SPDBs, distance, and delays between radio and UV/EUV LCs}
\label{sec_interp_delays}

In Fig.~\ref{fig_lc} we show that the peaks in the 1190 MHz flux record coincide with those on the LCs of some flare kernels within several seconds or tens of seconds. The agreement is not exact, but it is typically within the finite cadence of AIA. Considering that there are definite distances between radio sources and locations of flare kernels, the found temporal coincidences indicate that they could be caused by particle beams propagating between the 1190 MHz source (sources) and the flare kernels. For example, the velocity of beams generating the classical type III bursts was estimated as 0.15$c$ -- 0.6$c$ by \citet{Reid2018}, where $c$ is the speed of light. The beams with such velocities can explain the  time coincidences found.
We note that not all such beams can generate the radio emission. In particular, for the radio emission, the conditions for a two-step plasma emission process need to be fulfilled. That is, the beams need to form a bump on the distribution tail and the bump-on-tail instability, followed by transformation of Langmuir waves into electromagnetic waves \citep{1980gbs..bookR....M} for the radio emission to be produced.

As the observed SPDBs have positive frequency drift, the corresponding beams propagate downwards in the gravitationally stratified atmosphere. Taking as an example one burst of the SPDBs-2 at about 07:01 UT, with the maximum on 600 MHz at 07:01:08 UT and on 1100 MHz at 07:01:12 UT (see short white   lines in Fig.~\ref{r2}), the beam propagation lasts about 4 s. The corresponding maximum possible distance travelled by this beam is located along the longest loop above the HFT near N3, which is the loop connecting K3 and K13. Assuming that this loop is circular and has its top above K8, the maximum distance $L$ of beam propagation is $L \approx \pi/2 \times D$, where $D$ is the distance between K8 and K3 or K8 and K13, which in our case is about $\approx$ 150\arcsec. We thus obtain $L$\,$\lesssim$ 171\,Mm and a maximum beam propagation velocity of $\lesssim$ 42.8 Mm s$^{-1}$. This velocity is lower than the minimum velocity of the type III burst beams (0.15\,c = 45 Mm s$^{-1}$). For shorter loops, the velocity of the beams generating SPDBs would be even lower.

However, if the SPDBs  propagated along helical magnetic fields \citep[as found in previous studies by][]{Karlicky2002,Zemanova2020}, the beam propagation velocities could be higher on account of longer propagation lengths along their twisted trajectories \citep[see Fig. 7 of][]{Karlicky2002}. In models with a null-point, helical magnetic field lines were shown to be generated during the fan-spine reconnection \citep{Wyper2016a,Wyper2016b,Doyle2019}, which would support this idea. 
Therefore, we tried to identify helical structures along the longest newly created flare loops similar to those shown in Fig. 9d of \cite{Doyle2019}. However, measuring the twist is difficult here due to the overlapping of many flare loops
(Fig. \ref{fig_flare_1}g). The loops seem to be partially wrapped around one another, but even this may not bring enough twist to significantly alter the loop lengths,  and thus the beam propagation velocities.

The swirl we observed is likely related to the reconnection at the HFT since it occurs in its vicinity at the time when K9--K13 loops of S reconnect to form other large-scale flare loops (Fig. \ref{fig_schema}b and c, see Sect. \ref{sec_interp_evol}). It is an apparently rotating structure that likely contains evolving twisted field lines. Its relation to the HFT reconnection means that twist transfer to large-scale loops is possible (just as in the models of \citealt{Wyper2016a,Wyper2016b,Doyle2019}). Again, the amount of twist within the evolving swirl is difficult to measure, as there are many structures that cannot be resolved.

Another possibility is that our broad-band 600--1300 MHz SPDBs are not generated by a single beam propagating in the whole SPDB frequency range. It could be that SPDBs are a superposition of several shorter and narrower SPDBs generated in a cluster of loops with different plasma density, and thus in a broader frequency range than that of individual SPDBs. We regard this scenario as less likely since it involves multiple observationally unresolvable structures.
 
Be that as it may, in all cases, the angle between the beam propagation direction and the gradient of plasma density in gravitationally stratified solar atmosphere is expected to be high.

\subsection{Relation of kernel KA to the flare}
\label{sec_interp_KA}

Finally, we would like to comment on the observed enhancement of EUV intensity in loops of the neighbouring active region NOAA 12\,055 (Fig.~\ref{fig_12055}a, see also  Sects.~\ref{sec_flare} and~\ref{sec_uv_lc}). The bright kernel KA was partially synchronized in time with radio bursts (Fig.~\ref{fig_lc_12055}). This partial synchronization of LCs from the kernel in another active region could be explained by a magnetic linkage between these two active regions (Fig.~\ref{fig_12055}b). Considering that the peaks of UV LCs from kernels K13 and KA coincide within a few seconds with the peak of the radio flux at 1190 MHz, as well as the relatively large distances between those flare kernels, we suggest that only particle beams travelling along the magnetic field lines could be the agents explaining these observations.

Interestingly,  kernel KA is not located at the footpoints of 193 \AA~loops connecting regions 12\,055 and 12\,056. Rather, it is located on a branch of a system of quasi-separatrices \citep{Demoulin1995, Demoulin1996} 
that also extends to the footpoints of the  interconnecting loops. 
Therefore, we propose that kernel KA was not produced by a direct bombardment of the dense atmospheric layers by the particle beam propagating from the active region NOAA 12\,056. It is instead possible that the propagating particle beam could trigger a local energy release in the assumed current-carrying system around  kernel KA (for this mechanism, see \citealt{Karlicky1989}). 

%

\section{Summary}
\label{sec_summary}

We reported on the radio burst activity including rare slowly positively drifting bursts (SPDBs) during the SPDB-rich C7.8 flare of 2014 May 10, as well as the association of these bursts with the changes in magnetic connectivity observed in EUV.
During the flare, radio burst activity was observed during its impulsive phase and lasted until the beginning of the gradual phase. At 600--2000 MHz, we observed three groups of slowly positively drifting bursts (SPDBs-1 to 3) and one group of narrow-band type III bursts seen over the  continuum. Furthermore, within the 2000-5000 MHz range we observed continuum emission plus two groups consisting of several narrow-band type III bursts (groups III-A and III-B). The group SPDBs-1, as well as both III-A and III-B occurred during the peak of HXR flux at 25--50 keV. The other two groups of SPDBs occurred later, as the HXR flux was already declining.

An analysis of the imaging data from SDO/AIA revealed that the flare started when growing S 
started to interact with the nearby half dome structure. HXR sources at 25--50 keV were seen only along parallel ribbons. After the interaction of the S with the half dome started, we observed an EUV swirl in the vicinity of the S elbow, as well as the appearance of multiple kernels far south of the active region, followed by the appearance of large-scale hot EUV loops. We constructed 1600 \AA~UV and 304 \AA~EUV LCs in individual flare kernels, and compared them to the radio flux at 1190 MHz, at which frequency all of the SPDBs1--3 bursts were observed, as well as a group of narrow-band type III bursts. We found that especially  kernel K13 showed LCs that were in good agreement with the evolution of radio flux. This kernel corresponds to the footpoint of both the growing S and later the large-scale hot loops. The SPDBs are then a result of large-scale magnetic reconnection at HFT associated with the half dome, and their radio emission is generated by particle beams travelling along the reconnecting structures. In addition, at much higher frequencies, 2000--5000 MHz, we observed groups III-A and III-B of narrow-band type III bursts. 
These were found to be associated with the formation of the flare arcade and plasma densities of the order of 10$^{11}$\,cm$^{-3}$.

Moreover, we also found a similar response of UV/EUV emission to the  temporal evolution of radio flux at 1190 MHz  in  kernel KA located in the neighbouring active region. This time correlation (within seconds) between the radio flux at 1190 MHz and UV/EUV LCs, together with the large distance between kernels K13 and KA, imply that the agents synchronizing radio and UV/EUV LCs from this kernel are particle beams. 

Given that the SPDBs are produced by particle beams propagating downwards in the solar atmosphere, their low frequency drift can be explained by three different scenarios:
\begin{enumerate}
    \item The beam velocity is $\leq$ 0.15\,c, where 0.15\,c is the minimum velocity of beams generating type III bursts.
    \item The beam velocity is faster than 0.15\,c, but the beams move along the loop with helical magnetic field structure as found in previous works \citep{Karlicky2002,Zemanova2020}. Helical beam trajectories reduce the apparent distance of beam propagation along the visible loop \citep{Karlicky2002}. This idea is also supported by appearance of helical structures in the numerical models by \citet{Wyper2016a} and \citet{Wyper2016b}.
    However, in present observations, we did not identify any helical structures except some loops partially wrapping around one another.
    \item Another possibility is that these broad-band 600--1300 MHz SPDBs are not generated by a single beam propagating in the whole SPDB frequency range. Then SPDBs can be superposition of several shorter and narrower SPDBs. 
\end{enumerate} 
Of the three scenarios, we regard the first and second as the most likely, as the present radio and EUV observations, even if detailed, do not allow the  resolution of the individual hypothetical SPDBs. In both cases, the SPDBs are a result of large-scale magnetic reconnections in QSLs forming the HFT, and the angle between the beam propagation direction and the gradient of plasma density in a gravitationally stratified solar atmosphere is expected to be high.

\begin{acknowledgement}
We thank the referee for comments which helped to improve the manuscript. A.Z., M.K., J.D. and J.K. acknowledge institutional support from project RVO-67985815. A.Z., M.K., and J.K. also acknowledge support from the project GA\v{C}R grant 22-34841S. A.Z. and J.D. also acknowledge the support of GA\v{C}R grant 22-07155S and 20-07908S. 
J.R. acknowledges support from the Science Grant Agency project VEGA 2/0043/24 (Slovakia).
AIA and HMI data were provided courtesy of NASA/SDO and the AIA and HMI science teams. 
H-alpha data were provided by the Kanzelhöhe Observatory for Solar and Environmental Research, University of Graz, Austria. 
We acknowledge the use of the RHESSI Mission Archive available at https://hesperia.gsfc.nasa.gov/rhessi/mission-archive as well as the CALLISTO data by The Institute for Data Science FHNW Brugg/Windisch (Switzerland) available at https://www.e-callisto.org/Data/data.html.
\end{acknowledgement}

\bibliographystyle{aa} 
\bibliography{spdb_flare_20140510_references}      

\end{document}